\documentclass[notitlepage,a4paper,aps,prd,superscriptaddress,nofootinbib,groupedaddress]{revtex4}

\usepackage{enumerate}
\usepackage{amsmath}
\usepackage{amsfonts}
\usepackage{amssymb}
\usepackage[utf8]{inputenc}
\usepackage[T1]{fontenc}
\usepackage{mathtools}
\usepackage{wasysym}
\usepackage{accents}
\usepackage{graphicx}
\usepackage[english]{babel}
\usepackage[svgnames]{xcolor}
\usepackage[colorlinks=true,citecolor=blue]{hyperref}
\usepackage{url}
\usepackage{amsmath}
\usepackage{float}
\usepackage{mhchem}
\usepackage{bbold}
\usepackage{subcaption}

\newcommand{\JP}{
\affiliation{Physics Department, Federal University of Para\'iba, Caixa Postal 5008, 58059-900, Jo\~ao Pessoa-PB, Brazil}
}
\newcommand{\CG}{
\affiliation{Physics Department, Federal University of Campina Grande, Caixa Postal 10071, 58429-900, Campina Grande-PB, Brazil}
}
\newcommand{\UFCA}{
\affiliation{Instítuto de Formação de Educadores, Universidade Federal do Carir\'i, R. Olegário Emidio de Araujo, s/n - Centro, Brejo Santo - CE, 63260-000, Brazil}
}

\newcommand{\Khazar}{
\affiliation{Center for Theoretical Physics, Khazar University, 41 Mehseti Street, 1096 Baku, Azerbaijan}
}

\begin{document}

\title{Fermi Acceleration Mechanisms Beyond Lorentz Symmetry}

\author{Gilson A. Ferreira}\email{gilson.alves@estudante.ufcg.edu.br}
\CG

\author{Erick Aguiar}\email{erick.leite@academico.ufpb.br}
\JP

\author{A. A. Ara\'{u}jo Filho}\email{dilto@fisica.ufc.br}
\JP
\CG
\Khazar

\author{Edson Otoniel}\email{edson.otoniel@ufca.edu.br}
\UFCA

\author{Valdir B. Bezerra}\email{valdir@fisica.ufpb.br}
\JP

\author{Iarley P. Lobo}
\email{lobofisica@gmail.com,\\  iarley.lobo@academico.ufpb.br}
\JP
\CG

\begin{abstract}
   We construct models for first- and second-order Fermi acceleration of particles, incorporating generic frame transformations, dispersion relations, and conservation laws. Within this framework, we study deformations of Lorentz symmetry via the $\kappa$-Poincaré algebra in the bicrossproduct and classical bases, which respectively deform and preserve the relativistic dispersion relation. We also examine explicit Lorentz symmetry violation and compare the results with deformed relativity and special relativity. The energy spectra present different shapes when one considers deformation or violation of Lorentz symmetry in superluminal or subluminal scenarios. One of the possible outcomes is an intense decay of the spectrum for higher energies. We also gather the predictions of all the scenarios considered, comparing the spectra produced at the source in the energy domain, which is the variable in which cosmic ray observatories report their measurements.
\end{abstract}
\date{\today}
\maketitle
{
  \hypersetup{linkcolor=blue}
  \tableofcontents
}
%%%%%%%%%%%%%%%%%%%%%%%%%%%%%%%%%%%%%%%%%%%%%%%%%%%%%%%%%%%%%%%%%%%%%%%%%%%%%%%%%%%%%%%%%%%%%%%%%%%%%%%%%%%%%%%%%%%%%%%%%%%%%%%%%%%%%%%%%%%%%%%%%%%%%%%

\section{Introduction}

Lorentz symmetry is a consequence of the postulates of special relativity and represents one of the cornerstones of modern physics. It is a fundamental ingredient of the standard model of particle physics and is valid locally in the theory of general relativity. Despite its success, approaches aiming to quantize the gravitational field indicate that it may not be valid at a fundamental level. For instance, indications in this direction come from various quantum gravity theories, such as non-critical Liouville string theory \cite{Polchinski:1998rq,Mavromatos:2007xe}, loop quantum gravity \cite{Ashtekar:2021kfp,Amelino-Camelia:2016gfx}, causal dynamical triangulation \cite{Loll:2019rdj}, Horava-Lifshitz gravity \cite{Horava:2009if}, and 3D quantum gravity \cite{Freidel:2005me}, among others.

If Lorentz symmetry is not fundamentally valid, there are two main possibilities: either the local equivalence between inertial reference frames is no longer valid, meaning each observer assigns a different quantum gravity scale, or this equivalence is preserved via a different set of frame transformations that leave the scale invariant. The first case is referred to as the Lorentz Invariance Violation (LIV) scenario \cite{Mattingly:2005re}, while the second is known as Deformed Special Relativity (DSR) \cite{Amelino-Camelia:2000stu,Magueijo:2001cr}.

From a theoretical perspective, there are several ways to model the deformation of Lorentz invariance. For example, one can consider the invariance of the Hamiltonian in Finsler \cite{Girelli:2006fw,Amelino-Camelia:2014rga,Lobo:2016xzq,Pfeifer:2019wus,Lobo:2020qoa,Amelino-Camelia:2025ask} and Hamilton geometries \cite{Barcaroli:2015xda,Barcaroli:2017gvg} (for a review on the use of these geometries in quantum gravity, see \cite{Albuquerque:2023icp}), or through the structure of quantum algebras \cite{Lukierski:1991pn,Lukierski:1993wxa,Majid:1994cy,Borowiec:2009vb,Arzano:2021scz}. The latter approach has been widely studied by the quantum gravity community, with the $\kappa$-Poincar\'{e} algebra playing a prominent role as it describes the fundamental symmetry of the non-commutative $\kappa$-Minkowski spacetime \cite{Majid:1994cy,Pachol:2011tp}. The quantum gravity scale allows for different bases of this algebra, leading to different physical predictions. For instance, the so-called \textit{bicrossproduct basis} \cite{Majid:1994cy} features modified expressions for its Casimir operator (which defines the dispersion relation), inertial frame transformations, and energy-momentum conservation laws. Another important basis is the \textit{classical basis} \cite{Borowiec:2009vb,Carmona:2025fdu}, in which the Casimir operator and Lorentz transformations are preserved, but deformations emerge in the composition of energy and momentum and the action of symmetries on two-particle systems.

Some of the best phenomenological opportunities to observe quantum spacetime effects arise in scenarios where Lorentz symmetry is deformed or violated. Astrophysics has been a prolific source of observables and constraints with Planck-scale sensitivity. This means that even effects suppressed by powers of the Planck scale can produce detectable signatures due to the presence of amplifiers, such as the high energies of astroparticles and the vast distances they travel. For a review and roadmap on this topic, we recommend \cite{Addazi:2021xuf,AlvesBatista:2023wqm}; for a broader perspective on spacetime phenomenology in the ultraviolet and infrared regimes, see \cite{Amelino-Camelia:2008aez}.

Since the early days of quantum gravity phenomenology, cosmic rays have played a prominent role due to the ultra-high energies they can achieve (see, for instance, the discussion on quantum gravity effects on the GZK limit \cite{Amelino-Camelia:2001com} and its present status \cite{PierreAuger:2016use,PierreAuger:2023ankle,PierreAuger:2024origin}). Other effects include modifications to vacuum Cherenkov radiation \cite{Klinkhamer:2007ak}, the GZK photon spectrum \cite{Jacobson:2002hd}, and the development of air showers in both hadronic and electromagnetic sectors \cite{Boncioli:2015cqa,Diaz:2016dpk,PierreAuger:2021mve,Martinez-Huerta:2020cut}. In particular, the Pierre Auger Observatory has derived some of the most stringent constraints on Lorentz invariance violation in the propagation of ultra-high-energy cosmic rays and of the photons they produce \cite{PierreAuger:2022liv}. While most studies focus on propagation and detection effects, issues related to the sources of cosmic rays and their acceleration mechanisms are less explored.

One of the main mechanisms of acceleration of particles has been proposed by Fermi in 1949 \cite{Fermi:1949ee} and later refined to account for a better description of the physical environments in which acceleration can occur \cite{1978MNRAS.182..147B,1978MNRAS.182..443B,longair2011high}. They are based on the repeated interactions of particles with the environment that can be systematic (first order mechanism) or stochastic (second order mechanism) and, in general, involve some main ingredients: 1) first or second order in velocity transformation of energy and momentum between frames; 2) energy-momentum conservation law; 3) dispersion relation. The reason why these mechanisms are so successful lie in the fact that assuming special relativity kinematics, one derives an energy gain that allows an exponential growth of particles energy and a power law spectrum that matches observations \cite{Spurio:2018knn}.

Recently, important particle acceleration processes, known as first- and second-order Fermi acceleration mechanisms, were studied from the perspective of Lorentz Invariance Violation \cite{Duarte:2024aff}, where the effects of a modified dispersion relation (MDR) of the accelerated particles were considered. The impact of energy losses due to synchrotron radiation in this context was analyzed in \cite{Duarte:2025lnw}. Besides the dispersion relation, other crucial ingredients for Fermi mechanisms are the transformation rules between frames and the conservation laws of energy and momentum, which determine the energy gain in each cycle as relativistic particles interact with their environment. This leads to a natural question: \textit{what is the impact of other kinematical deformations of relativity on Fermi mechanisms?}

In this paper, we construct a general framework to analyze Fermi mechanisms when not only modified dispersion relations are involved, but also modified Lorentz transformations and energy-momentum conservation laws. Specifically, we apply this framework to the $\kappa$-Poincar\'{e} algebra in the bicrossproduct basis, which incorporates all these features, and compare it with the case of Lorentz violation, where only the MDR is considered. We also consider a case that provides a unique signature of the DSR scenario by examining the classical basis of the $\kappa$-Poincar\'{e} algebra. This allows us to investigate quantum gravity effects on Fermi acceleration without an MDR and with standard Lorentz transformations, but with a modified energy-momentum composition law.

Besides the ingredients highlighted above, a complete characterization of acceleration mechanisms should also involve the environment in which the particles are accelerated through electromagnetic, thermodynamic and magnetohydrodynamics effects, among others \cite{longair2011high}. However, we should analyze extra contributions, how, and if, they are important in another opportunity. One of the reasons for this choice is the lack of a rigorous field theoretical and thermodynamical description of the kind of deformed relativity scenario that we are addressing. Besides that, it is possible that extra contributions are suppressed in comparison to relativistic results, for instance, the equation of state of the gas that interacts with the particles in the first order mechanism is based on a non-relativistic treatment, therefore quantum gravity corrections of this treatment would be negligible \cite{Spurio:2018knn}. Nevertheless, we shall adopt a pragmatic approach of considering at least phenomenological corrections of the kinematics of accelerated particles as a first step towards a more complete characterization of quantum gravity effects on acceleration mechanisms.

The spectra that emerge from this framework are naturally written in terms of the dimensionless combination $x=\ell E$, which is the variable that best exhibits the dependence on the deformation or violation of the symmetry. Cosmic ray observatories, on the other hand, report their measurements as functions of the energy itself~\cite{PierreAuger:2020ehi}. For this reason, we devote a section of the paper to gathering the predictions of all the scenarios that we analyze and to comparing them, at the level of the spectrum produced at the source, in the energy domain, once a value of the deformation scale is fixed. We emphasize that this is a comparison among the models, and not a comparison with data. A confrontation with the measured flux would additionally require the injected mass composition, the propagation of nuclei through the cosmic radiation backgrounds, the distribution and evolution of the sources, magnetic deflections, and the detector response, all of which are essential ingredients of the combined analyses performed by the collaborations~\cite{PierreAuger:2023ankle,PierreAuger:2024origin} and lie beyond the scope of the kinematical study presented here.

The paper is organized as follows. In Sections~\ref{sec:fermi_gen} and \ref{sec:2ndorder}, we revisit the first- and second-order Fermi acceleration mechanisms for general frame transformations, dispersion relations, and energy-momentum conservation laws (such revision is important for identifying where the main kinematical hypotheses of the mechanisms lie). In Section~\ref{sec:kappa}, we review the main results from the $\kappa$-Poincar\'{e} algebra in the bicrossproduct and classical bases, and define how we model Lorentz violation using a parameter that controls the deformation. In Sections~\ref{sec:first-beyond} and \ref{sec:2nd-beyond}, we apply these results to the first- and second-order Fermi mechanisms and calculate the spectrum and spectral index via a diffusion equation. In Section~\ref{sec:compare}, we collect the source spectra predicted by the different models and compare them in the energy domain. Finally, in Section~\ref{sec:disc}, we summarize and discuss the main results of the paper.

%%%%%%%%%%%%%%%%%%%%%%%%%%%%%%%%%%%%%%%%%%%%%%%%%%%%%%%%%%%%%%%%%%%%%%%%%%%%%%%%%%%%%%%%%%%%%%%%%%%%%%%%%%%%%%%%%%%%%%%%%%%%%%%%%%%%%%%%%%%%%%%%%%%%%%%

\section{First order Fermi acceleration mechanism for a general dispersion relation, Lorentz transformation and composition law}\label{sec:fermi_gen}

In this section, we revisit the first-order Fermi acceleration mechanism within a general framework, without specifying the form of the frame transformations, the energy-momentum conservation law and the dispersion relation. This approach will serve as a basis for studying cases involving the deformation and violation of symmetries later on. The revisiting of the fundamentals of the acceleration mechanisms is essential for identifying where their main kinematical hypotheses lie and how the Lorentz symmetry and the conservation laws play the significant role.

%%%%%%%%%%%%%%%%%%%%%%%%%%%%%%%%%%%%%%%%%%%%%%%%%%%%%%%%%%%%%%%%%%%%%%%%
\subsection{Diffusive shock acceleration}\label{sec:shock}
Since the 1970s, the dominant acceleration mechanism in astrophysics has been associated with strong shock waves, referred to as ``diffusive shock acceleration.'' Such shocks can be found in several scenarios, like in supernova remnants, pulsar wind nebula, active galactic nuclei jets, gamma-ray bursts, solar winds and mergers of galactic clusters \cite{longair2011high}.  Consider a shock wave that separates a gas into two regions: an upstream part that is unperturbed, with parameters ($p_1,\rho_1,T_1$), and a downstream part containing gas perturbed by the passage of the shock front, with parameters ($p_2,\rho_2,T_2$).

The shock front travels with velocity $v_s$ in the frame where the unperturbed gas (upstream region) is stationary. Suppose relativistic particles are isotropically distributed in the upstream region. These particles move with velocities much greater than the shock front's velocity and that of the perturbed gas, which moves with velocity $U=3v_s/4$ according to the thermodynamics of a non-relativistic gas \cite{Spurio:2018knn}. When a particle crosses the shock front and enters the downstream region, it suffers an elastic collision and is pushed back towards the upstream region. In the frame where the downstream region is at rest, these particles are also isotropically distributed and travel from a region moving with velocity $U$ in the opposite direction. Each time a particle crosses the discontinuity surface, it is elastically kicked from one frame to another moving with a relative velocity $U$ opposed to its direction. When this happens, the particle gains a percentage of energy $\eta$.

%%%%%%%%%%%%%%%%%%%%%%%%%%%%%%%%%%%%%%%%%%%%%%%%%%%%%%%%%%%%%%%%%%%%%%%%%%%%%%%%%%%%

\subsection{First order mechanism}\label{sec:fermi1}

Each time a particle crosses the shock wave, it moves from the downstream to the upstream frame \cite{Spurio:2018knn,longair2011high}. Consider a particle with energy $E$ and momentum $\vec{p}$ in frame $S$, and $E'$ and $\vec{p'}$ in frame $S'$. The frames move relative to each other with speed $U$. The energy-momentum of the particle in frame $S'$ is given by:
\begin{align}
    &E'=f_0(E,\vec{p},U),\\
    &\vec{p'}=\vec{f}(E,\vec{p},U),
\end{align}
where $f_0$ and $\vec{f}$ constitute the transformation between frames $S$ and $S'$. These are the Lorentz transformations in special relativity or in a Lorentz-violating scenario, but they could take a different form in a deformed relativistic framework, such as the $\kappa$-Poincar\'{e} algebra discussed in this paper. We also stress that considering these frame transformations, just the first-order term in $U/c$ is relevant, thus we discard terms of order ${\cal O}(U^2/c^2)$ or higher. Only the infinitesimal transformation is sufficient. For this reason, we call it the first-order mechanism.

The energy of the particle in frame $S$ is found by inverting this transformation:
\begin{align}
  &E=f_0(E',\vec{p'},-U),\label{eq:transf-e}\\
    &\vec{p}=\vec{f}(E',\vec{p'},-U).\label{eq:transf-p}  
\end{align}

When we consider an elastic collision between the particle and the region where the cloud is located, the energy and momentum obey the following relation:
\begin{equation}\label{eq:antip0}
    E'\mapsto E', \qquad \vec{p'}\mapsto \ominus \vec{p'}.
\end{equation}

In this expression, we assume that momentum conservation could differ from the usual conservation law given by the sum of momenta. In principle, it could be given by a deformed law, represented by the symbol $\oplus$, such that the composition of the 4-momenta $p$ and $q$ is given by $p\oplus q$. We refer to $\ominus p$ as the momentum that annihilates $p$, i.e., the one for which $p\oplus (\ominus p)=0$. We express the conservation equation in this general framework because it is also present in deformed relativistic scenarios \cite{Gubitosi:2011hgc}.

Therefore, after an elastic collision of the particle with the cloud, the energy in frame $S$ transforms to $E^*$ as:
\begin{align}
    &E=f_0(E',\vec{p'},-U)\mapsto E^*=f_0(E',\ominus \vec{p'},-U),
\end{align}
which, using Eqs.~\eqref{eq:transf-e} and \eqref{eq:transf-p}, leads to the following energy after the collision:
\begin{equation}
    E^*=f_0\left(f_0(E,\vec{p},U),\ominus \vec{f}(E,\vec{p},U),-U\right).\label{eq:gained-en}
\end{equation}

It is possible to avoid the dependence of $E^*$ on $\vec{p}$ by writing the velocity of the particle as:
\begin{equation}\label{eq:vel1}
    v^i=\frac{\partial E}{\partial p_i}\Bigg|_{C=m^2c^4},
\end{equation}
and expressing $\vec{p}$ as a function of $v^i$, $E$, and $m$, where $C$ is the on-shell relation. It is possible to simplify \eqref{eq:gained-en} further by recalling that we are considering relativistic particles, such that the norm of their velocity is approximately given by the speed of light $c_{\ell}$. This consists of taking the approximation $m\approx 0$. We use the notation $c_{\ell}$ to express that the speed of light can differ from $c$, depending on the form of the dispersion relation, and can be modified by the action of an inverse energy scale we call $\ell$ (which is would be the quantum gravity scale).

The quantity
\begin{equation}\label{eq:en-gain}
   \Delta E= E^*-E
\end{equation}
measures the energy gain in each collision. By averaging over all directions, we derive the average energy gain and the efficiency of the mechanism, which represents the percentage of energy gain in each passage between frames:
\begin{equation}
    \eta=\left\langle\frac{\Delta E}{E}\right\rangle.
\end{equation}

In summary, we can find the energy gain $\Delta E$ from \eqref{eq:en-gain} and \eqref{eq:gained-en} for any relativistic framework, given the following expressions:
\begin{enumerate}
    \item the infinitesimal transformations $f_0$ and $\vec{f}$ that act on energy and momentum from \eqref{eq:transf-e} and \eqref{eq:transf-p},
    \item the opposite momentum $\ominus \vec{p}$ from \eqref{eq:antip0},
    \item the deformed velocity of particles $v^i$ from \eqref{eq:vel1}.
\end{enumerate}

Any of these ingredients can produce a deformed energy gain and modify the acceleration process.

%%%%%%%%%%%%%%%%%%%%%%%%%%%%%%%%%%%%%%%%%%%%%%%%%%%%%%%%%%%%%%%%%%%%%%%%%%%%%%%%%%%%%%%%%%%%%%%%%%%%
\subsubsection{Particle spectrum}
From Bell's argument \cite{1978MNRAS.182..147B,1978MNRAS.182..443B}, we can derive the particle spectrum using a diffusion-loss equation. The number density of particles $N$ at a given time $t$ within an energy range $(E,E+\mathrm{d}E)$ is given by:
\begin{equation}
    \frac{\mathrm{d}N}{\mathrm{d}t}=D\nabla^2N+Q-\frac{N}{\tau_{\mathrm{esc}}}+\frac{\partial(b\cdot N)}{\partial E},
\end{equation}
where $D$ is a diffusion coefficient, $Q$ is the rate of particle injection per unit volume, $b$ is the energy loss rate, and $\tau_{\mathrm{esc}}$ is the time a particle remains within the acceleration region, which we call the escape time (see section 7.5.1 of \cite{longair2011high}). Since we are interested in a steady-state solution and neglect sources, $\mathrm{d}N/\mathrm{d}t=\nabla^2N=Q=0$, and we are left with:
\begin{equation}\label{eq:diff}
  \frac{N}{\tau_{\mathrm{esc}}}-  \frac{\partial(b\cdot N)}{\partial E}=0.
\end{equation}

The energy gain can be described by the average energy gain $\langle \Delta E\rangle$ and the average time between particle collisions, which is the characteristic time scale of the process, as \cite{Spurio:2018knn,Duarte:2024aff}:
\begin{equation}
    b=-\frac{\mathrm{d}E}{\mathrm{d}t}=-\frac{\langle \Delta E\rangle}{\tau_{\mathrm{avg}}}.   
\end{equation}
We also have that $\tau_{\mathrm{esc}}$ and $\tau_{\mathrm{avg}}$ are related by:
\begin{equation}
    \tau_{\mathrm{esc}}=\frac{\tau_{\mathrm{avg}}}{1-P},
\end{equation}
where $P$ is the probability that the particle remains in the acceleration region, called the residence probability, and $P_{\mathrm{esc}}=1-P$ is the escape probability \cite{Spurio:2018knn,Duarte:2024aff}. Part of the expression for $b$ we already have, since we know $\langle \Delta E\rangle$. However, we need to derive an expression for $P$. This is where we use Bell's argument \cite{1978MNRAS.182..147B,1978MNRAS.182..443B}.

%%%%%%%%%%%%%%%%%%%%%%%%%%%%%%%%%%%%%%%%%%%%%%%%%%%%%%%%%

\subsubsection{The escape probability}

The flux is the same in either direction of particle passage, meaning that at each pass, the particle gains energy \cite{Spurio:2018knn}. In the downstream region, due to the random velocity of the particles, described by the isotropic nature of the velocity distribution in the rest frame of the downstream region, there is a probability that the particles leave the shock region and are lost. The escape probability is the probability that, after a collision, a particle escapes the acceleration region and is lost for future iterations. 

The differential number density of isotropic particles of velocity $v$ is given by $\mathrm{d}n/\mathrm{d}E=4\pi\Phi(E)/v$, where $\Phi(E)$ is the differential flux of particles of a given energy in a given solid angle. Since we are describing a planar surface (as we are talking about planar shock waves in \ref{sec:shock}), the differential particle flux with energy $E$ through a planar surface is given by ${\cal F}(E)=\pi \Phi(E)$ \cite{Spurio:2018knn}. This gives the differential flux for relativistic particles, with $v\approx c_{\ell}$, through the shock front:
\begin{equation}
    {\cal F}(E)=\frac{c_{\ell}}{4}\frac{\mathrm{d}n}{\mathrm{d}E},
\end{equation}
where the factor $4$ comes from the geometrical shape of the ``detector'' surface.

In the frame in which the shock front is at rest, the gas in the upstream region moves towards the front at velocity $v_1=v_s$, and the gas (assumed as monatomic) in the downstream region moves away from the front at a velocity $v_2=v_s/4$ (from the thermodynamics of a non-relativistic monoatomic gas) \cite{longair2011high,Spurio:2018knn}. Therefore the differential rate in which particles can be removed from the shock region is given by differential flux of particles at speed $v_s/4$ towards the boundary of the acceleration region:
\begin{equation}
    {\cal F}_{\mathrm{esc}}= v_2\frac{\mathrm{d}n}{\mathrm{d}E}=\frac{v_s}{4}\frac{\mathrm{d}n}{\mathrm{d}E}.
\end{equation}

The fraction of particles lost per unit time is given by the ratio between the flux of particles escaping from the acceleration region and the flux of particles crossing the planar front. Therefore, the probability that a particle is removed from the acceleration region is given by:
\begin{equation}
    P_{\mathrm{esc}}=\frac{{\cal F}_{\mathrm{esc}}}{{\cal F}(E)}=\frac{v_s}{c_{\ell}}.
\end{equation}

In the original Bell argument, the integral flux of particles is used instead of the differential one. However, in this case, the integral flux would not be proportional to the number density of particles, since the speed of the high-energy particle is possibly energy-dependent beyond Lorentz symmetry. This makes the energy dependence of the escape process explicit. For this reason, we consider the differential flux instead.

%%%%%%%%%%%%%%%%%%%%%%%%%%%%%%%%%%%%%%%%%%%%%%%%%%%%%%%%%

\subsubsection{The spectral index}

Since the velocity of the shock front is given by $v_s=4U/3$, where $U$ is the velocity of the perturbed gas in the downstream region in the rest frame of the unperturbed gas in the upstream region, we derive that the escape probability is related to the residence probability $P$ as:
\begin{equation}
    1-P=P_{\mathrm{esc}}=\frac{4}{3}\frac{U}{c_{\ell}}.
\end{equation}

Using these expressions in \eqref{eq:diff}, we derive:
\begin{align}
   &\frac{N}{\tau_{\mathrm{avg}}}(1-P)+\frac{1}{\tau_{\mathrm{avg}}}\left(N\frac{\mathrm{d}\langle \Delta E\rangle}{\mathrm{d}E}+\langle \Delta E\rangle\frac{\mathrm{d}N}{\mathrm{d}E}\right)=0\\
   \Rightarrow & \qquad \langle \Delta E\rangle\frac{\mathrm{d}N}{\mathrm{d}E}+N\left(\frac{\mathrm{d}\langle \Delta E\rangle}{\mathrm{d}E}+\frac{4}{3}\frac{U}{c_{\ell}}\right)=0.\label{eq:spectrum}
\end{align}

This means that we have, formally, the same kind of equation as in special relativity. The difference lies in the speed of light term, which can be a function of energy, and in the energy dependence of $\langle \Delta E\rangle$, which can be non-linear.

An energy-dependent spectral index $\lambda$ can be found from $N(E)=N_0 E^{\lambda(E)}$ or from:
\begin{equation}\label{eq:spec0}
    \lambda(E)=\frac{\mathrm{d} \log(N(E))}{\mathrm{d} \log (E)}=\frac{E}{N(E)}\frac{\mathrm{d}N(E)}{\mathrm{d}}E.
\end{equation}

If we define a new variable $x=\ell E$, where $\ell=\kappa^{-1}$ is the inverse of the energy scale $\kappa$, the spectral index is given by:
\begin{equation}\label{eq:spec1}
    \lambda(x)=\frac{x}{N(x)}\frac{\mathrm{d}N(x)}{\mathrm{d}x}.
\end{equation}

%%%%%%%%%%%%%%%%%%%%%%%%%%%%%%%%%%%%%%%%%%%%%
\section{Second order Fermi acceleration mechanism for a general dispersion relation, Lorentz transformation and composition law}\label{sec:2ndorder}

Fermi's second-order mechanism was originally proposed in 1949 \cite{Fermi:1949ee} and was based on the idea of generating acceleration through a stochastic process by which particles collide with the interstellar medium. The original idea modeled charged particles being reflected from ``magnetic mirrors,'' which are irregularities in the galactic magnetic field that move randomly. The main difference between Fermi's original argument and today's formulation of the second order acceleration consists in the fact that the particle's energy changes all the time in a stochastic way, which means that in the calculation, we have to consider the statistical nature of the acceleration mechanism, as well as the systematic increase in energy (already described in the first order case).

The collision between the particle and the mirror is such that the angle between the initial direction of the particle and the normal to the surface of the magnetic mirror is $\theta$. The mirror is considered infinitely massive, and its velocity is not modified in the collision. Let us consider the center of momentum reference frame to be that of the cloud (frame $S'$) moving with velocity $U$ relative to the observer frame $S$.

The analysis is very similar to the first-order mechanism discussed in section \ref{sec:fermi1}. The only difference is that we shall consider the Lorentz transformation up to second order in $U$ (for this reason, it is called the second-order mechanism). If we write the energy-momentum of the particle in the cloud frame, we have:
\begin{align}
    &E'=g(E,\vec{p},U),\\
    &\vec{p}'=\vec{g}(E,\vec{p},U),
\end{align}
where $(g,\vec{g})$ are the Lorentz transformations up to second order in $U$. The energy gain reads formally the same as in the first-order case; however, we consider the frame transformation up to second order in $U/c$ \eqref{eq:gained-en}:
\begin{equation}\label{eq:estar2}
    E^*=g_0\left(g_0(E,\vec{p},U),\ominus \vec{g}(E,\vec{p},U),-U\right),
\end{equation}
where the dependence on $\vec{p}$ vanishes when we write it as a function of the particle's velocity \eqref{eq:vel1}. The quantity $\Delta E= E^*-E$ measures the energy gain in each collision.

Because of the random scattering, we need to average $\Delta E$ over the pitch angle $\theta$. However, we must consider the probability $\mathrm{Pr}(\theta)$ of a collision occurring at a certain angle to average this quantity. We assume that the probability of the angle lying in the interval $(\theta,\theta+\mathrm{d}\theta)$ is proportional to $\sin(\theta)\, \mathrm{d}\theta$, and we average over the full range $(0,\pi)$.

Randomness is the key element that distinguishes this mechanism from the first-order case. Thus, the average energy gain after one cycle of acceleration is given by:
\begin{equation}\label{eq:aver2}
    \langle \Delta E\rangle=\frac{\int_0^{\pi} \mathrm{Pr}(\theta) \Delta E(\theta)\sin(\theta)\, \mathrm{d}\theta}{\int_0^{\pi}\mathrm{Pr}}(\theta)\sin(\theta)\,\mathrm{d}\theta,
\end{equation}
where $\mathrm{Pr}(\theta)$ is the probability of encounters taking place at an angle $\theta$. This probability is proportional to the rate of encounters of high-energy particles with the accelerating obstacle in the laboratory reference frame. At first order in $U/c$, it is proportional to the time difference between the arrival of particles in the rest frame of the obstacle plus a relativistic factor due to the relative distance these particles propagate before reaching the obstacle at the speed of light $c_{\ell}$ (see, for instance, section 9.3 of \cite{longair2011high}):

\begin{equation}\label{eq:prob_bicross}
    P(\theta)\propto 1+\frac{U}{c_{\ell}}\cos(\theta)+{\cal O}(U^2/c_{\ell}^2).
\end{equation}

We should stress that we can keep the probability at first order in $U/c_{\ell}$ because the term $\Delta E$ is already given by the sum of a quantity proportional to $U$ and another proportional to $U^2$.

This average energy is important for calculating the spectrum, as we will see below. For a particle that remains in the accelerating region for a time $\tau_{\mathrm{esc}}$, we have a diffusion-loss equation similar to the first-order mechanism. However, considering the statistical nature of the acceleration process, as well as the systematic increase in energy, we need to consider the Fokker-Planck equation for diffusion in momentum space \cite{longair2011high} (see Eq.(7.42) and (17.22). This equation can be written as a diffusion-loss equation with an extra term (the last one in the expression below):
\begin{equation}
    \frac{\mathrm{d}N}{\mathrm{d}t}=D\nabla^2N+Q-\frac{N}{\tau_{\mathrm{esc}}}+\frac{\partial(b(E)\cdot N)}{\partial E}+\frac{1}{2}\frac{\partial^2}{\partial^2 E}\left[d(E)\cdot N\right],
\end{equation}
where, as before, $b(E)=-\mathrm{d}E/\mathrm{d}t$, and the new contribution $d(E)=\mathrm{d}\langle (\Delta E)^2\rangle/\mathrm{d}t$ is the mean square change in energy per unit time.

If the mean free path between clouds is $L$, the time between collisions of relativistic particles with speed $c_{\ell}$ is given by $L/(c_{\ell}\cos(\phi))$, where $\phi$ is the angle between the direction of the particle and the magnetic field direction. Averaging $\cos(\phi)$, we find that the average time between collisions is $\tau_{\mathrm{col}}=2L/c_{\ell}$. In special relativity, the term $b(E)$ would be proportional to the energy $E$, with the proportionality factor given by $\alpha=4U^2/(3cL)$. For this reason, let us redefine the quantity $L$ in terms of the new quantity $\alpha$ to make it easier to compare with special relativity.

Thus, we can write:
\begin{align}
   &b(E)=-\frac{\mathrm{d}E}{\mathrm{d}t}=-\frac{\langle \Delta E\rangle}{\tau_{\mathrm{col}}}=-\frac{3c\alpha}{8U^2}c_{\ell}\langle \Delta E\rangle,\label{eq:b2}\\
   &d(E)=\frac{\mathrm{d}\langle(\Delta E)^2\rangle}{\mathrm{d}t}=\frac{\langle(\Delta E)^2\rangle}{\tau_{\mathrm{col}}}=\frac{3c\alpha}{8U^2}c_{\ell}\langle(\Delta E)^2\rangle.\label{eq:d}
\end{align}

We also consider the stationary case in which $\mathrm{d}N/\mathrm{d}t=D\nabla^2 N=Q=0$, which gives:
\begin{equation}\label{eq:fokker}
     \frac{N}{\tau_{\mathrm{esc}}}-\frac{\partial(b(E)\cdot N)}{\partial E}-\frac{1}{2}\frac{\partial^2}{\partial^2 E}\left[d(E)\cdot N\right]=0.
\end{equation}

In special relativity, as described in \cite{longair2011high}, this approach gives a power-law spectrum of the form $N_{\mathrm{SR}}\propto E^{-y}$, where $y$ is a model-dependent parameter that depends on $\alpha$ and $\tau_{\mathrm{esc}}$. To match the result $y=-2$, one must have:
\begin{equation}\label{eq:tauesc}
\tau_{\mathrm{esc}}=\alpha^{-1}.    
\end{equation}

For this reason, let us also use this convention so that we can compare the results of departures from Lorentz symmetry with the special relativistic result. Thus, our final equation is:
\begin{equation}
   N+\frac{3c}{8U^2}\frac{\partial(c_{\ell}\langle \Delta E\rangle\cdot N)}{\partial E}-\frac{3c}{16U^2}\frac{\partial^2}{\partial^2 E}\left[c_{\ell}\langle(\Delta E)^2\rangle\cdot N\right]=0.
\end{equation}

Again, from the solution of this equation, we can define the spectral index as in Eqs.~\eqref{eq:spec0} and \eqref{eq:spec1}.

%%%%%%%%%%%%%%%%%%%%%%%%%%%%%%%%%%%%%%%%%%%%%%%%%%%%%%%%%%%%%%%%%%%%%%%%%%%%%%%%%%%%%%%%%%%%%%%%%%%%%%%%%%%%%%%%%%%%%%%%%%%%%%%%%%%%%%%%%%%%%%%%%%%%%%%

\section{Deformation and Violation of Lorentz Symmetry}\label{sec:kappa}

We review two relevant bases of the $\kappa$-Poincar\'{e} algebra. The first is the bicrossproduct basis, which contains a modified dispersion relation, deformed Poincar\'{e} symmetry for the one-particle sector, and a modified composition law \cite{Majid:1994cy}. The second is the so-called classical basis, which presents the standard dispersion relation and Poincar\'{e} transformations in the one-particle sector but contains deformations in the two-particle sector, modifying the conservation law of energy and momentum and the action of symmetries on two-particle momenta \cite{Pachol:2011tp}.

%%%%%%%%%%%%%%%%%%%%%%%%%%%%%%%%%%%%%%%%%%%%%%%%%%%%%%%%%%%%%%%%%%%%%%%%%%%%%
\subsection{Bicrossproduct basis}

The deformed classical algebra of symmetry generators is given by \cite{Gubitosi:2011hgc} ($\ell=\kappa^{-1}$ is a deformation parameter with dimensions of inverse energy):
\begin{align}
   &\{P_{\mu},P_{\nu}\}=0,\qquad \{N_j,E\}=P_j,\nonumber \\
    &\{N_j,P_k\}=\delta_{jk}\left(\frac{1}{2c^2\ell}(1-e^{-2\ell P_0})+\frac{\ell}{2}|\vec{P}|^2\right)-\ell P_jP_k,\nonumber\\
    &\{N_j,N_k\}=-\epsilon_{jkl}\frac{R_l}{c^2},\qquad \{R_j,E\}=0,\qquad \{R_j,P_k\}=\epsilon_{jkl}P_l,\nonumber \\
    &\{R_j,N_k\}=\epsilon_{jkl}N_l,\qquad \{R_j,R_k\}=\epsilon_{jkl}R_l,
\end{align}
where $P_{\mu}$ are the generators of spacetime translations (momenta), $|\vec{P}|^2=\sum_iP_i$, $N_{i}$ is the boost generator in the $i$-direction, and $R_{i}$ generate each of the three rotations. We highlight that this algebra is usually presented in units where $c=1$. To restore the speed of light scale, we transformed $(N_i,P_j)\mapsto (cN_i,cP_j)$ on both sides of the algebra relations \cite{Gubitosi:2011hgc}. The mass Casimir of this algebra is:
\begin{equation}
    C=\frac{4}{\ell^2}\sinh^2\left(\frac{\ell}{2}P_0\right)-c^2e^{\ell P_0}|\vec{P}|^2,
\end{equation}
where the on-shell relation is given by $C=m^4c^2$. In the limit $\ell\rightarrow 0$, we recover the Poincar\'{e} algebra.

To study these effects on momenta, we use canonical coordinates with Poisson brackets that satisfy:
\begin{align}
    \{t,E\}=1,\qquad \{x^i,p_j\}=\delta^i_j, 
\end{align}
and zero otherwise. A representation of energy, $x$-direction momentum, $x$-direction boost generators, and mass Casimir in these canonical coordinates is given by:
\begin{align}
    &P_0=E,\qquad P_1=p_x,\qquad N_x=tp_x+x\left(\frac{1-e^{-2 \ell E}}{2 c^2\ell}+\frac{\ell}{2} p^2-\ell p_x^2\right)\label{eq:gen1}\\
   &C=\frac{4}{\ell^2}\sinh^2\left(\frac{\ell}{2}E\right)-c^2e^{\ell E}p^2,\label{eq:casimir}\, .
\end{align}
where we define $p^2=|\vec{p}|^2$. The infinitesimal boost between momenta in the $x$-direction is given by $p'_{\mu}=p_{\mu}+U\{N_x,p_{\mu}\}$:
\begin{align}
    &E'=E+Up_x,\label{eq:booste1}\\
    &p_x'=p_x+U\left(\frac{1-e^{-2 \ell E}}{2 c^2\ell}+\frac{\ell}{2} p^2-\ell p_x^2\right),\label{eq:boostp1}
\end{align}
where we can restrict to the analysis on the $x$-direction, since we are going to make a later average over different angles, which will be sufficient for our purposes.

To guarantee the invariance of the nature of interactions, we require a covariance condition with a modified composition of momenta as:
\begin{equation}
    \Lambda(p\oplus q)=\Lambda(p)\oplus \Lambda(q), 
\end{equation}
where $\Lambda(p)_{\mu}=p_{\mu}'$ is the boosted momentum.\footnote{Usually, it is necessary to require a backreaction on the boost parameter when applied to the ``second'' particle in the conservation law, as described in \cite{Gubitosi:2011hgc,Lobo:2021yem}, but we omit these details here, since they are unnecessary for our purposes.} From an algebraic point of view, this is realized by using the coproduct structure accommodated in the $\kappa$-Poincar\'{e} Hopf algebra. In this coordinate system, the composed energy and momentum of particles $1$ and $2$ read \cite{Gubitosi:2011hgc,Amelino-Camelia:2024tdk}:
\begin{align}
    &E_1\oplus E_2=E_1+E_2,\\
    &p_{i,1}\oplus p_{i,2}=p_{i,1}+e^{-\ell E_1}p_{i,2}.
\end{align}

The conservation law is fundamental for describing particle interactions \cite{Lobo:2021yem} and will be relevant when we describe elastic interactions that accelerate particles. One consequence of the deformed composition law is that the opposite momentum $\ominus p_{\mu}$ that annihilates a given one ($p_{\mu}\oplus(\ominus p_{\mu})=0$) is given by:
\begin{align}
&\ominus E=-E,\\
&\ominus p_i=-e^{\ell E}p_i.\label{eq:antip1}
\end{align}

\subsubsection{Momentum as a function of velocity and the speed of light}\label{sec:speed}

To conclude the brief review of the main results of the bicrossproduct basis of the $\kappa$-Poincar\'{e} algebra, we also derive the speed of particles from the modified dispersion relation defined by the mass Casimir. From \eqref{eq:casimir}, we can find the energy and its first derivative:
\begin{align}
&E=\ell^{-1}\log \left(\frac{2}{2+c^4 m^2 \ell ^2-c \ell  \sqrt{c^6 m^4 \ell ^2+4 \left(c^2 m^2+p^2\right)}}\right)\label{eq:e1}\\
    &\frac{\partial E}{\partial p_x}=\frac{4 c p_x}{\sqrt{c^6 m^4 \ell ^2+4 \left(c^2 m^2+p^2\right)} \left[c^4 m^2 \ell ^2-c \ell  \sqrt{c^6 m^4 \ell ^2+4 \left(c^2 m^2+p^2\right)}+2\right]}
\end{align}

Imposing the on-shell condition $C=m^2c^4$ from \eqref{eq:casimir}, we find the speed of particles in the $x$-direction:
\begin{equation}\label{eq:vx1}
    v_x=\frac{\partial E}{\partial p_x}\Bigg|_{C=m^2c^4}=\frac{4 c^2 \ell  e^{2 E \ell } p_x}{\sqrt{\left(e^{2 E \ell } \left(c^2 p^2 \ell ^2-1\right)+1\right)^2} \left[1-e^{2 E \ell } \left(c^2 p^2 \ell ^2-1\right)-\sqrt{\left(e^{2 E \ell } \left(c^2 p^2 \ell ^2-1\right)+1\right)^2}\right]}
\end{equation}

It is interesting to solve this equation as a function of the momentum $p_x$. To do this, we consider the MDR \eqref{eq:casimir}, which gives:
\begin{equation}\label{eq:p2}
    p^2=\frac{e^{-\ell E} \left(4 \sinh ^2\left(\ell E/2\right)-c^4 m^2 \ell ^2\right)}{c^2 \ell ^2}.
\end{equation}

By substituting \eqref{eq:p2} back into \eqref{eq:vx1}, and using the property that $v_x=v\cos(\theta)$ (where $v^2=|\vec{v}|^2$ and $\theta$ is the projection angle along the $x$-axis), we can solve \eqref{eq:vx1} for $p_x$ as:
\begin{equation}
    p_x=\frac{v e^{-2\ell E} \cos (\theta ) \left[e^{\ell E} \left(c^4 m^2 \ell ^2+2\right)-2\right]}{2 c^2 \ell }.
\end{equation}

A corollary of these computations is that if we compute $v_y$ and $v_z$ from Eq.~\eqref{eq:vx1} by mapping $x\mapsto y$ and $x\mapsto z$, and consider $m=0$, we can find the speed of massless particles as an energy-dependent quantity:
\begin{equation}\label{eq:ckappa}
    c_{\ell}(E)=\sqrt{v_x^2+v_y^2+v_z^2}\Bigg|_{m=0}=ce^{\ell E},
\end{equation}
which means that when $\ell$ is positive/negative, we are in a superluminal/subluminal scenario.

For relativistic particles, we can approximate $v\approx c_{\ell}$, which gives us an expression for the momentum in the $x$-direction as:
\begin{equation}\label{eq:pxrel}
    p_x\approx \frac{1-e^{-\ell E}}{ c \ell }\cos (\theta ).
\end{equation}

%%%%%%%%%%%%%%%%%%%%%%%%%%%%%%%%%%%%%%%%%%%%%%%%%%%%%%%%%%%%%%%%%%%%%%%%%%%%%%%%%%%%%%%%%%%
\subsubsection{Modeling Lorentz Violation}

The set of conditions that describe a deformed relativistic scenario is given by the deformed algebra of generators (which defines the mass shell) and the composition law. Without any of these ingredients, we are in a Lorentz-violating scenario. The quantum gravity community usually assumes such a scenario as one produced by a modification of the dispersion relation without the other ingredients, implying an on-shell relation given by $C=m^2c^4$ from \eqref{eq:casimir}, but keeping the standard Lorentz transformations between frames and the usual composition law. We can model both the deformation and violation of Lorentz symmetry in this paper by introducing a parameter $\epsilon$, such that the Lorentz transformation and composition law are modified by a parameter $\epsilon \ell$ instead of simply $\ell$. When $\epsilon=1$, we are in the $\kappa$-Poincar\'{e} scenario, and when $\epsilon=0$, we are in a Lorentz-violating scenario.

Besides the dispersion relation \eqref{eq:casimir}, the main equations we will need in this work are the boosts in the $x$-direction \eqref{eq:booste1} and \eqref{eq:boostp1} and the antipode in the $x$-direction \eqref{eq:antip1}, which can be written as:
\begin{align}
    &f_0(E,\vec{p},U)=E'=E+Up_x,\label{eq:f0-bi}\\
    &f_x(E,\vec{p},U)=p_x+U\left(\frac{1-e^{-2 \epsilon\ell E}}{2 c^2\epsilon\ell}+\frac{\epsilon\ell}{2} p^2-\epsilon\ell p_x^2\right),\label{eq:f1-bi}\\
    &\ominus p_x=-e^{\epsilon\ell E}p_x\label{eq:om-bi}
\end{align}
where we introduced the functions $f_0$ and $f_1$ to describe the action of a boost on energy and the $x$-momentum, respectively. We are preserving the usual deformed Casimir operator \eqref{eq:casimir} and the expressions from subsection \ref{sec:speed} from \eqref{eq:e1} to \eqref{eq:pxrel}.

%%%%%%%%%%%%%%%%%%%%%%%%%%%%%%%%%%%%%%%%%%%%%%%%%%%%%%%%%%%%%%%%%%%%%%%%%%%%%%%%%%%%%%%%%%%%%%%%%%%%%%%%%%%%%%%%%%%%%%%%%%%%%%%%%%%%%%%%%%%%%%%%%%%%%%%
\subsection{Classical basis}

Besides the bicrossproduct case, a basis that has gained attention recently is one that preserves the dispersion relation and Lorentz transformations of special relativity and introduces modifications in the composition law and the action of symmetries on composed momenta: the classical basis. Interesting aspects of this basis include the fact that constraints on quantum gravity from in-vacuo dispersion \cite{Vasileiou:2013vra,LHAASO:2024lub} do not apply to it because massless particles have speed $c$, nor do threshold effects bounds (as in the bicrossproduct case) because the tight constraints based on particle interactions assume Lorentz violation \cite{HAWC:2019gui}. Thus, it is possible that the symmetries of special relativity are modified by a scale larger than the Planck energy without contradicting any experimental result so far, offering rich phenomenological opportunities \cite{Carmona:2025fdu}.

The modified composition law of the classical basis is the following:
\begin{align}
    &E_1\oplus E_2=E_1\Pi(p_2)+\frac{1}{\Pi(p_1)}\left(E_2+\ell c^2\vec{p}_1\cdot \vec{p}_2\right)\\
    &p_{i,1}\oplus p_{i,2}=p_{i,1}\Pi(p_2)+p_{i,2},
\end{align}
where 
\begin{equation}
    \Pi(p)=\ell E+\sqrt{1+\ell^2\left(E^2-c^2p^2\right)}\stackrel{\text{on-shell}}{=}\ell E+\sqrt{1+\ell^2c^4m^2}.
\end{equation}

In this case, the only expression we will need for our computations is the representation of the antipode action as:
\begin{equation}
    \ominus \vec{p'}=-\vec{p}\left(\ell E+\sqrt{1+\ell^2c^4m^2}\right).\label{eq:om-cl}
\end{equation}

%%%%%%%%%%%%%%%%%%%%%%%%%%%%%%%%%%%%%%%%%%%%%%%%%%%%%%%%%%%%%%%%%%%%%%%%%%%%%%%%%%%%%%%%%%%%%%%%%%%%%%%%%%%%%%%%%%%%%%%%%%%%%%%%%%%%%%%%%%%%%%%%%%%%%%%

\section{First order Fermi Mechanism beyond Lorentz symmetry}\label{sec:first-beyond}

Let us consider the case of deformed relativity with a modified dispersion relation given by the bicrossproduct basis of the $\kappa$-Poincar\'{e} algebra and the LIV scenario with only the modified dispersion relation. After that, we consider the classical basis, which represents a deformed relativistic scenario affecting interactions but not particle propagation, as described by the classical basis of the $\kappa$-Poincar\'{e} algebra.

%%%%%%%%%%%%%%%%%%%%%%%%%%%%%%%%%%%%%%%%%%%%%%%%%%%%%%

\subsection{The bicrossproduct basis of $\kappa$-Poincar\'{e} algebra and LIV scenarios}

Using the prescription described in section \ref{sec:fermi1} with the results of section \ref{sec:kappa} (for ultrarelativistic particles $m\approx 0$) and Eqs.~\eqref{eq:f0-bi}, \eqref{eq:f1-bi}, \eqref{eq:om-bi}, we obtain the energy variation as:
\begin{align}
    \Delta E=\frac{e^{-\ell E} \left(e^{ \ell E}-1\right)  \left(e^{\epsilon \ell  E}+1\right)}{c \ell }U\cos (\theta ).
\end{align}

In the DSR ($\epsilon=1$) and LIV ($\epsilon=0$) scenarios, we have:
\begin{align}
    &\Delta E_{\kappa,\mathrm{b}}=2 U \cos (\theta )\frac{ \sinh ( \ell E)}{c \ell },\quad
    &\Delta E_{\mathrm{LIV}}=2 U \cos (\theta )\frac{1- e^{-\ell E}}{c \ell },
\end{align}
where we use the subscript $\kappa,\mathrm{b}$ to denote the $\kappa$-Poincar\'{e} results in the bicrossproduct basis, and the subscript $\mathrm{LIV}$ for the Lorentz-violating case, where we consider only the effect of the modified dispersion relation without the other algebraic results.

We calculate the average energy deposition over all directions, since this is an isotropic process. Using the fact that $\langle\cos(\theta)\rangle=\frac{\int_0^1\cos(\theta)\cdot \cos(\theta)\,\mathrm{d}\cos(\theta)}{\int_0^1\cos(\theta)\,\mathrm{d}\cos(\theta)}=\frac{2}{3}$, we therefore have:
\begin{align}
   &\langle \Delta E\rangle_{\kappa,\mathrm{b}}=\frac{4}{3}\frac{U}{c} \frac{\sinh ( \ell E)}{\ell},\quad 
    &\langle \Delta E\rangle_{\mathrm{LIV}}=\frac{4}{3}\frac{U}{c} \frac{(1-e^{\ell E})}{\ell}.
\end{align}

The efficiency is measured as $\eta=\langle\Delta E/E\rangle$. For the $\kappa$-Poincar\'{e} and LIV scenarios, it is given by:
\begin{align}
    &\eta_{\kappa,\mathrm{b}}= \frac{4}{3}\frac{U}{c} \frac{\sinh ( \ell E)}{\ell E}\approx \frac{4}{3}\frac{U}{c} \left(1+\frac{\ell^2 E^2}{6}\right),\quad
    &\eta_{\mathrm{LIV}}=\frac{4}{3}\frac{U}{c} \frac{(1-e^{\ell E})}{\ell E}\approx \frac{4}{3}\frac{U}{c} \left(1-\frac{\ell E}{2}\right).
\end{align}

In the special relativity (SR) limit, the efficiency is given by $\eta_{\mathrm{SR}}=\frac{4}{3}\frac{U}{c}$, where we use the subscript $\mathrm{SR}$ to denote special relativistic quantities. In contrast to the special relativistic result, the efficiency in these scenarios is energy-dependent. The LIV case presents first-order corrections, while the $\kappa$-Poincar\'{e} case presents only much weaker corrections, given by second-order contributions in the quantum gravity scale. We see that for the LIV case, the mechanism is less/more efficient than the SR one in the superluminal/subluminal scenario. On the other hand, since $\eta_{\kappa,\mathrm{b}}$ is an even function of $\ell$, both cases produce a more efficient way of accelerating particles compared to SR.

%%%%%%%%%%%%%%%%%%%%%%%%%%%%%%%%%%%%%%%%%%%%%%%%%%%%%%%%%%%%%%%%%%%%%%%%%%
\subsubsection{Particle spectrum (super and subluminal cases)}

In the case of $\kappa$-Poincar\'{e} and LIV, we use \eqref{eq:spectrum} to find:
\begin{align}\label{eq:N1-k}
     N_{\kappa,\mathrm{b}}^{(1)}{}'(E)+ \ell \frac{e^{2\ell E}+3}{e^{2 \ell E}-1}N_{\kappa,\mathrm{b}}^{(1)}{}(E)=0,\quad N_{\mathrm{LIV}}^{(1)}{}'(E)+ \frac{2\ell}{e^{\ell E}-1}N_{\mathrm{LIV}}^{(1)}{}(E)=0,
\end{align}
where prime $'$ denotes differentiation with respect to energy, and we use the superscript $(1)$ to denote the first-order mechanism.

In Special Relativity, we have $c_{\ell}=c$ and $\langle \Delta E\rangle=4U/3c$, which gives:
\begin{equation}
  E\frac{\mathrm{d}N_{\mathrm{SR}}}{\mathrm{d}E}+2N_{\mathrm{SR}}=0\Rightarrow N_{\mathrm{SR}}(E)\propto E^{-2}, 
\end{equation}
with a spectral index of $-2$.

If we redefine the variables as $x=\ell E$, set $c=1$, and consider the case $\ell>0$ (the superluminal scenario, denoted by the subscript $+$), we can rewrite expressions \eqref{eq:N1-k} as:
\begin{align}
     &N_{\kappa,\mathrm{b},+}^{(1)}{}'(x)+ \frac{e^{2x}+3}{e^{2 x}-1}N_{\kappa,\mathrm{b},+}^{(1)}{}(x)=0,\quad N_{LIV,+}^{(1)}{}'(x)+ \frac{2}{e^{x}-1}N_{LIV,+}^{(1)}{}(x)=0.
\end{align}

We have an analytical solution for these expressions given by:
\begin{align}
    N_{\kappa,\mathrm{b},+}^{(1)}{}(x)\propto \frac{\text{csch}(x)}{4} \left[\coth (x)+1\right],\quad N_{LIV,+}^{(1)}{}(x)\propto \frac{e^{2 x}}{\left(e^x-1\right)^2}.
\end{align}

The spectral index \eqref{eq:spec1} can be found as:
\begin{align}
    \lambda_{\kappa,\mathrm{b},+}^{(1)}{}(x)=x\left[1-2\coth(x)\right],\quad \lambda_{LIV,+}^{(1)}{}(x)=-\frac{2 x}{e^x-1}.
\end{align}

\begin{figure}[H]
    \centering
    \begin{subfigure}{0.45\textwidth}
        \centering
        \includegraphics[width=\textwidth]{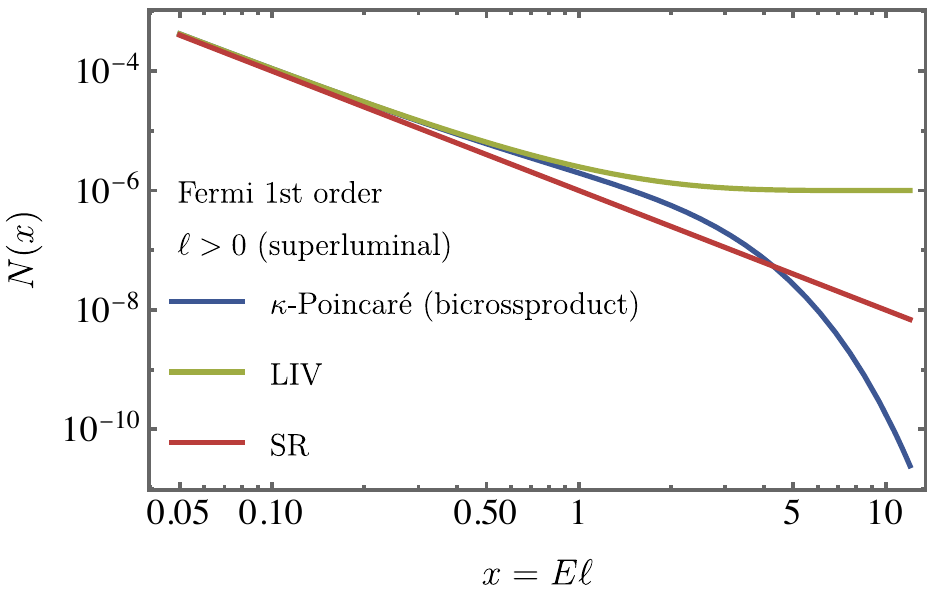}
        \caption{Spectra of Fermi first-order mechanism in the superluminal scenario for DSR (blue), LIV (green), and SR (red) cases.}
        \label{fig:spectrum-bicross-super}
    \end{subfigure}
    \hfill
    \begin{subfigure}{0.45\textwidth}
        \centering
        \includegraphics[width=\textwidth]{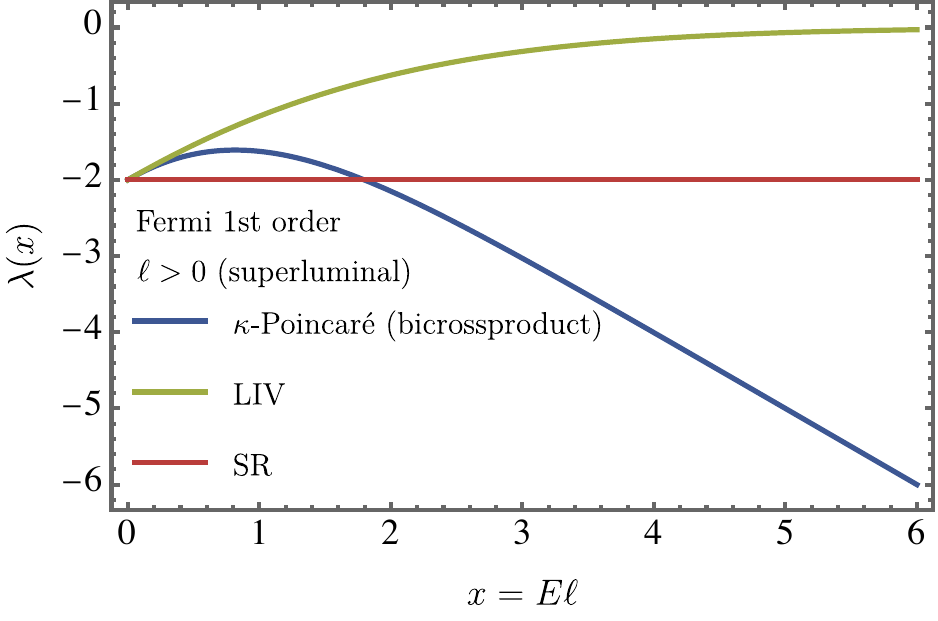}
        \caption{Spectral indexes of Fermi first-order mechanism in the superluminal scenario for DSR (blue), LIV (green), and SR (red) cases.}
        \label{fig:index-bicross-super}
    \end{subfigure}
    \caption{First-order Fermi mechanism in the superluminal scenario considering Lorentz preservation, violation, and deformation in the bicrossproduct basis of the $\kappa$-Poincar\'{e} algebra.}
    \label{fig:bic-super}
\end{figure}

As can be seen in Fig.~\ref{fig:spectrum-bicross-super}, in the superluminal case, the $\kappa$-Poincar\'{e} and LIV spectra decay at a slower rate than the SR case. However, around the deformation scale, the LIV case approaches a constant configuration, while the $\kappa$-Poincar\'{e} case drops at a stronger rate than the SR case. This behavior can also be seen in Fig.~\ref{fig:index-bicross-super}, which shows the spectral index.

The subluminal scenario can be analyzed for $\ell<0$, which can be achieved by transforming $\ell\mapsto -\ell$ in the diffusion equation. The equations and solutions are the same as in the previous case, with the transformation $x\mapsto -x$. The solutions are:
\begin{equation}
    N_{\kappa,\mathrm{b},-}^{(1)}{}(x)\propto \frac{\text{csch}(x)}{4} \left[\coth (x)-1\right],\quad N_{LIV,-}^{(1)}{}(x)\propto \frac{1}{\left(e^x-1\right)^2},
\end{equation}
and the spectral indexes read:
\begin{align}
    \lambda_{\kappa,\mathrm{b},-}^{(1)}{}(x)=-x\left[1+2\coth(x)\right],\quad \lambda_{LIV,-}^{(1)}{}(x)=-\frac{2x e^x}{e^x-1}.
\end{align}

\begin{figure}[H]
    \centering
    \begin{subfigure}{0.45\textwidth}
        \centering
        \includegraphics[width=\textwidth]{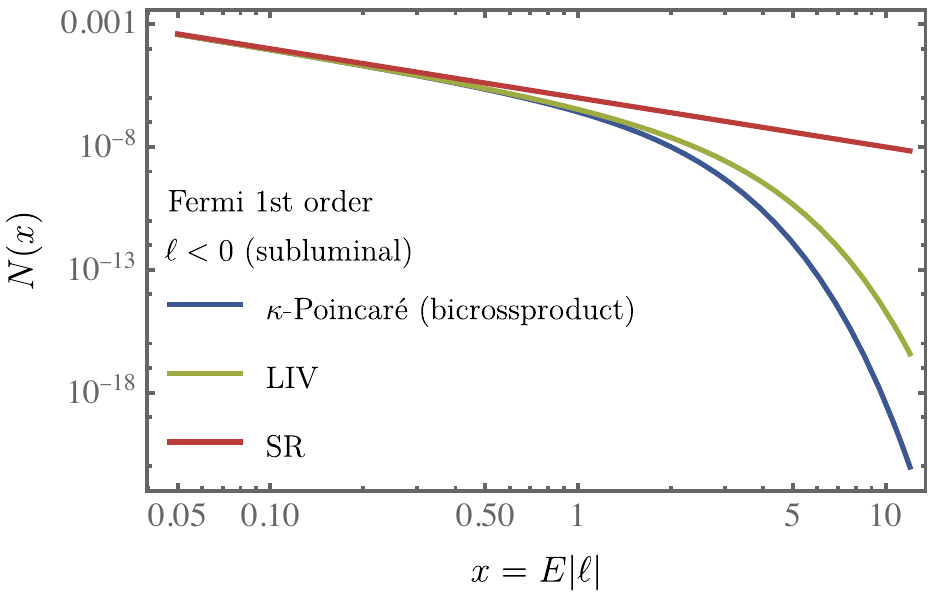}
        \caption{Spectra of Fermi first-order mechanism in the subluminal scenario for DSR (blue), LIV (green), and SR (red) cases.}
        \label{fig:spectrum-bicross-sub}
    \end{subfigure}
    \hfill
    \begin{subfigure}{0.45\textwidth}
        \centering
        \includegraphics[width=\textwidth]{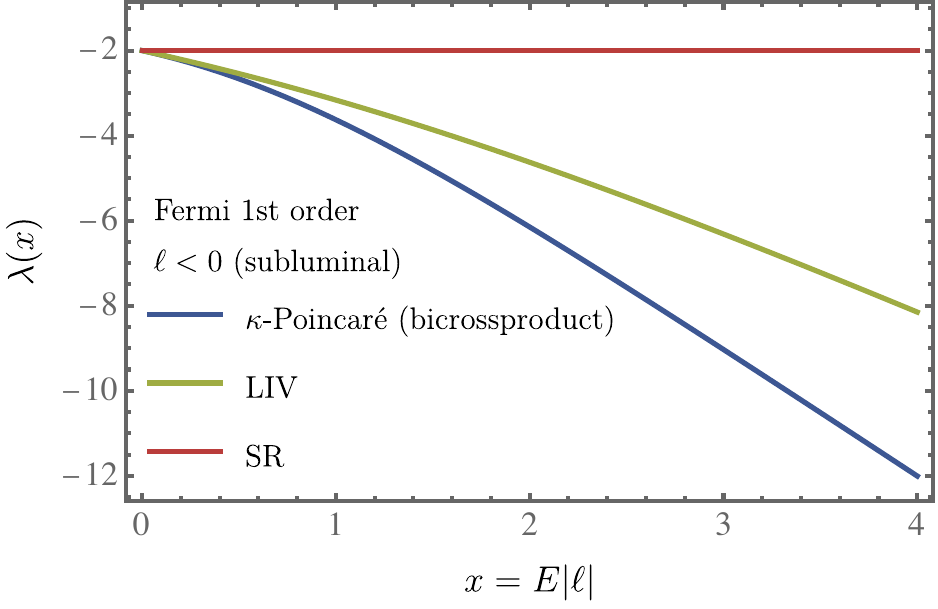}
        \caption{Spectral indexes of Fermi first-order mechanism in the subluminal scenario for DSR (blue), LIV (green), and SR (red) cases.}
        \label{fig:index-bicross-sub}
    \end{subfigure}
    \caption{First-order Fermi mechanism in the subluminal scenario considering Lorentz preservation, violation, and deformation in the bicrossproduct basis of the $\kappa$-Poincar\'{e} algebra.}
    \label{fig:bic-sub}
\end{figure}

As can be seen in Figs.~\ref{fig:spectrum-bicross-sub} and \ref{fig:index-bicross-sub}, in this scenario, the spectra decay monotonically in a more intense way than the SR counterpart, with a stronger effect in the DSR case than in the LIV case.

%%%%%%%%%%%%%%%%%%%%%%%%%%%%%%%%%%%%%%%%%%%%%%%%%%%%%%%%%%%%%%%%%%%%
\subsection{Classical basis}

Using the prescription described in section \ref{sec:fermi1} with the results of \ref{sec:kappa} (for ultrarelativistic particles $m\approx 0$) and Eq.~\eqref{eq:om-cl}, we obtain the energy variation as:
\begin{equation}
    \Delta E_{\mathrm{cl}}=\frac{U}{c} E\cos (\theta )\left(1+\frac{1 }{1+ \ell E}\right),
\end{equation}
where we use the subscript $\mathrm{cl}$ to denote the classical basis. The efficiency reads:
\begin{equation}
    \eta_{\mathrm{cl}}=\frac{4}{3}\frac{U}{c}\frac{1+\ell E/2}{1+ \ell E},
\end{equation}
which becomes less efficient with increasing energy, approaching $0.5 \eta_{\mathrm{SR}}$ when $E\gg \ell^{-1}$. In this case, massless particles follow the speed of light, and we do not have super- or subluminal scenarios. The spectral equation becomes (in terms of the variables $E$ and $x=\ell E$):
\begin{align}
   &E\left(\ell^2 E^2+3 \ell E+2\right) N_{\mathrm{cl}}^{(1)}{}'(E)+N_{\mathrm{cl}}^{(1)}{}(E) \left(3 \ell^2 E^2+6 \ell E +4\right)=0,\\
   &x \left(x^2+3 x+2\right) N_{\mathrm{cl}}^{(1)}{}'(x)+\left(3 x^2+6 x+4\right) N_{\mathrm{cl}}^{(1)}{}(x)=0,
\end{align}
respectively.

The solution of these equations can be found analytically:
\begin{align}
    &N_{\mathrm{cl}}^{(1)}{}(x)\propto \frac{x+1}{x^2 (x+2)^2},\\
    &\lambda_{\mathrm{cl}}^{(1)}{}(x)=\frac{4}{x+2}-\frac{1}{x+1}-3,
\end{align}
where we use the superscript $(1)$ to denote the first-order mechanism.

\begin{figure}[H]
    \centering
    \begin{subfigure}{0.45\textwidth}
        \centering
        \includegraphics[width=\textwidth]{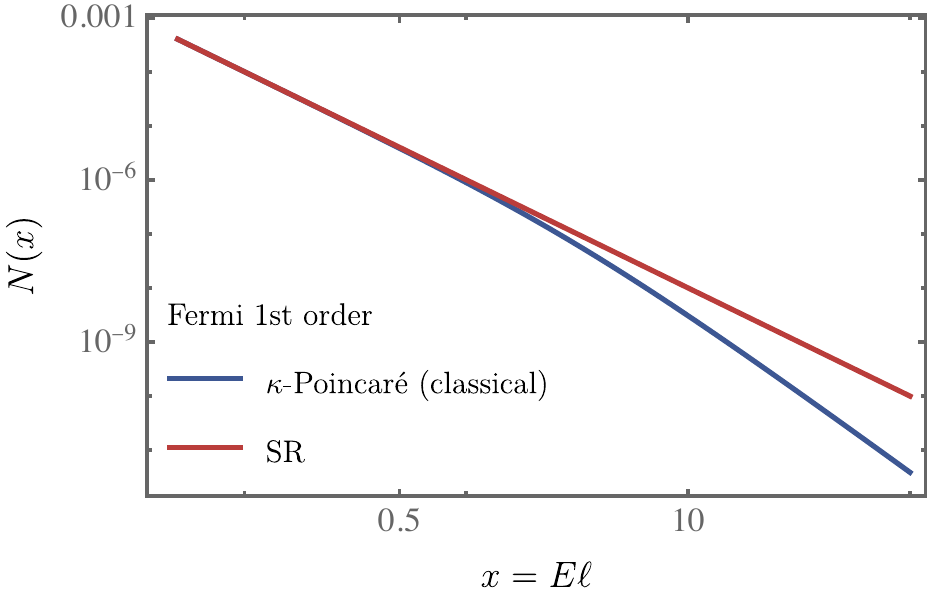}
        \caption{Spectra of Fermi first-order mechanism for DSR (blue) and SR (red) cases.}
        \label{fig:spectrum-class}
    \end{subfigure}
    \hfill
    \begin{subfigure}{0.45\textwidth}
        \centering
        \includegraphics[width=\textwidth]{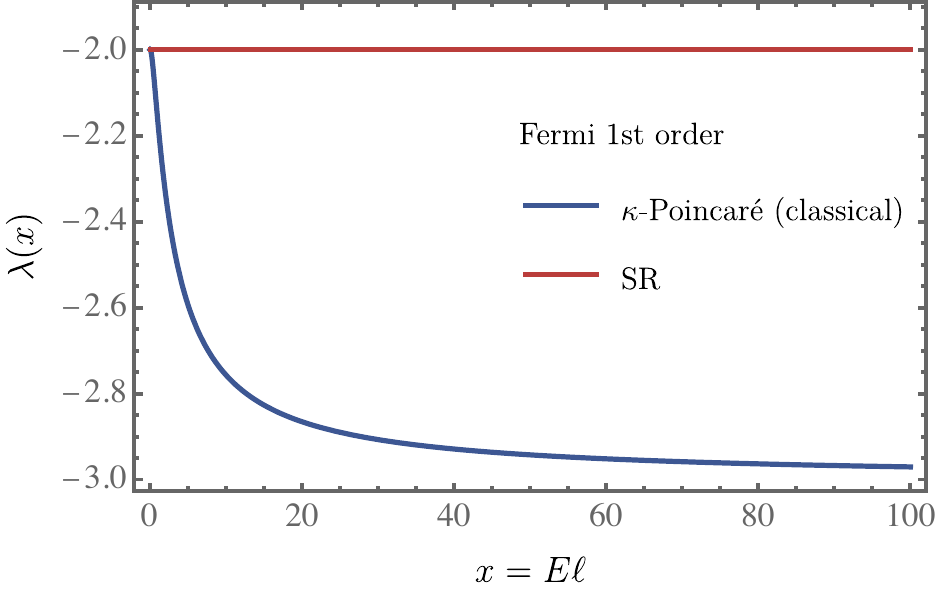}
        \caption{Spectral indexes of Fermi first-order mechanism for DSR (blue) and SR (red) cases.}
        \label{fig:index-class}
    \end{subfigure}
    \caption{First-order Fermi mechanism considering Lorentz preservation and deformation in the classical basis of the $\kappa$-Poincar\'{e} algebra.}
    \label{fig:first-class}
\end{figure}

As can be seen in Figs.~\ref{fig:spectrum-class} and \ref{fig:index-class}, the spectral index transitions from the value $-2$ to $-3$ at high energies.

\begin{figure}[H]
    \centering
    \includegraphics[width=0.5\linewidth]{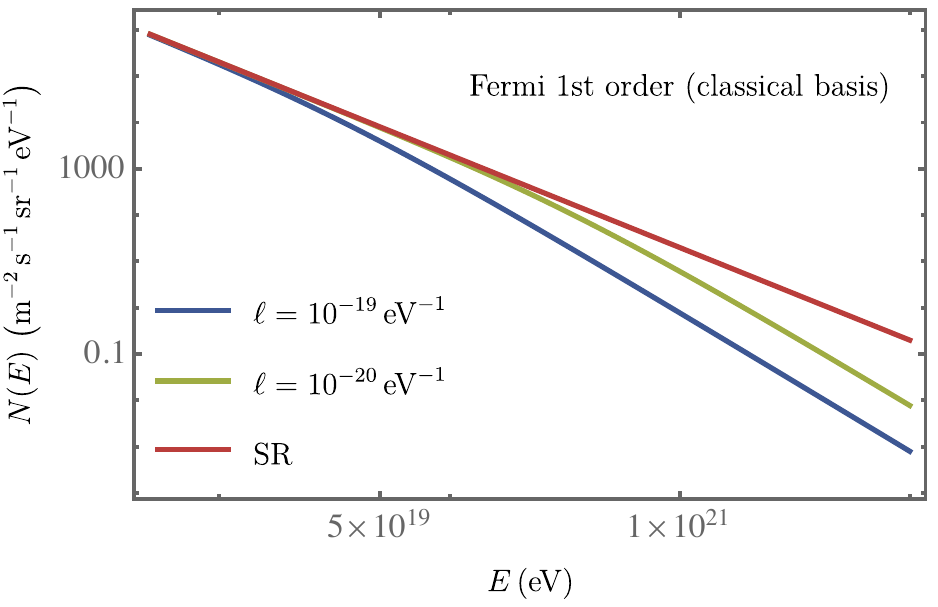}
    \caption{Spectrum, derived from Fermi 1st order mechanism for different deformation parameters of the classical basis of the $\kappa$-Poincar\'{e} algebra.}
    \label{fig:dnde}
\end{figure}

It is interesting to note that since this basis of the $\kappa$-Poincar\'{e} algebra does not present a modified dispersion relation, bounds placed from observables based on this modification do not apply, such as bounds from time delays of photons \cite{Vasileiou:2013vra,LHAASO:2024lub} or threshold effects in interactions \cite{HAWC:2019gui,PierreAuger:2022liv}. Therefore, an unknown physical effect with a scale smaller than the Planck scale could trigger a transition of acceleration effects from a spectral index of $-2$ to $-3$. To illustrate this situation, let us consider the effect in Fig. \ref{fig:dnde} for different deformation parameters. In this case, one observes that the inclination of the differential spectrum transitions from $-2$ to $-3$ for energies beyond the deformation energy scale.

%%%%%%%%%%%%%%%%%%%%%%%%%%%%%%%%%%%%%%%%%%%%%%%%%%%%%%%%%%%%%%%%%%%%%%%%%%%%%%%%%%%%%%%%%%%%%%%%%%%%%%%%%%%%%%%%%%%%%%%%%%%%%%%%%%%%%%%%%%%%%%%%%%%%%%%

\section{Second order Fermi Mechanism beyond Lorentz symmetry}\label{sec:2nd-beyond}

In this section, we analyze the effects of approaches that go beyond Lorentz symmetry on the second-order Fermi acceleration.

%%%%%%%%%%%%%%%%%%%%%%%%%%%%%%%%%%%%%%%%%%%%%%%%%%%%%%%%%%%%%%%%%%%%%%%%%%%%%%%%%%%%%%%%%%%%%%%%%

\subsection{The bicrossproduct basis of $\kappa$-Poincar\'{e} algebra and LIV scenarios}

At second order, the deformed Lorentz transformation in the $x$-direction is generated by the boost generator $N_x$ \eqref{eq:gen1} as:
\begin{equation}
    p_{\mu}'=p_{\mu}+U\{N_x,p_{\mu}\}+\frac{U^2}{2}\left\{N_x,\{N_x,p_{\mu}\}\right\}.
\end{equation}

This implies that the functions $g_{\mu}$ we shall consider are given by:
\begin{align}
    &g_0(E,\vec{p},U)=E+Up_x +\frac{U^2}{2} \left(\frac{1- e^{-2 \epsilon\ell E}}{2 c^2 \epsilon\ell}+\frac{\epsilon\ell p^2}{2}-\epsilon\ell p_x^2\right),\\
    &g_x(E,\vec{p},U)=p_x+U\left(\frac{1-e^{-2 \epsilon\ell E}}{2 c^2\epsilon\ell}+\frac{\epsilon\ell}{2} p^2-\epsilon\ell p_x^2\right)+\frac{U^2}{2}\left[\frac{e^{-2\ell E}p_x}{c^2}-\ell p_x\left(\frac{1-e^{-2 \epsilon\ell E}}{2 c^2\epsilon\ell}+\frac{\epsilon\ell}{2} p^2-\epsilon\ell p_x^2\right)\right],
\end{align}
where, as before, we use the parameter $\epsilon$ to control whether we are considering a deformation of Lorentz symmetry using the bicrossproduct basis of the $\kappa$-Poincar\'{e} algebra ($\epsilon=1$) or using only the modified dispersion relation with standard Lorentz boost transformations ($\epsilon=0$).

In presenting the results, we distinguish between the DSR scenario, where we use the subscripts $\kappa,\mathrm{b}$ to indicate that the result is derived from $\epsilon=1$ and describes the bicrossproduct basis of the $\kappa$-Poincar\'{e} algebra, and the subscript $\mathrm{LIV}$ to describe the Lorentz Violating Scenario. We also use the superscript $(2)$ to denote the second-order Fermi mechanism. From this, we can use the result of \eqref{eq:estar2} to find the energy gain:

\begin{align}
    \Delta E_{\kappa,\mathrm{b}}^{(2)}&=\frac{U e^{-\ell E } \left(e^{\ell E }-1\right) \left(e^{\ell E }+1\right) \cos (\theta )}{c \ell }-\frac{U^2 e^{-2 \ell E } \left(e^{\ell E }-1\right) \left(\left(e^{\ell E }-1\right) \left(e^{2 \ell E }+1\right) \cos ^2(\theta )-2 e^{\ell E } \left(e^{\ell E }+1\right)\right)}{2 c^2 \ell }\\
    &\approx U^2 \left(\frac{2 E}{c^2}-\frac{E^2 \ell  \cos ^2(\theta )}{c^2}\right)+\frac{2 E U \cos (\theta )}{c},\nonumber\\
   \Delta E_{\mathrm{LIV}}^{(2)}&=\frac{2 E U^2}{c^2}+\frac{2 U \cos (\theta ) \left(1- e^{-\ell E }\right)}{c \ell }\approx \frac{2 E U^2}{c^2}+U \left(\frac{2 E \cos (\theta )}{c}-\frac{E^2 \ell  \cos (\theta )}{c}\right).
\end{align}

To calculate the average energy gain, we need to average this result over the pitch angles and consider the probability of an encounter taking place at an angle $\theta$, given by Eq.~\eqref{eq:prob_bicross}:

\begin{equation}
    P(\theta)\propto 1+\frac{U}{c}e^{-\ell E}\cos(\theta).
\end{equation}

Using equation \eqref{eq:aver2}, the average energy in the DSR and LIV scenarios is given by:
\begin{align}
    \langle\Delta E\rangle_{\kappa,\mathrm{b}}^{(2)}&=-\frac{U^2 e^{-2 \ell E } \left(e^{\ell E }-1\right) \left(-5 \ell E  e^{\ell E }+\ell E  e^{3 \ell E }-\ell E -e^{2 \ell E } (7 \ell E +2)+2\right)}{6 c^2 \ell E ^2}\nonumber\\
    &\approx \frac{8 E U^2}{3 c^2}-\frac{2 \ell  \left(E^2 U^2\right)}{3 c^2}+\frac{5 E^3 U^2 \ell ^2}{9 c^2},\\
    \langle\Delta E\rangle_{\mathrm{LIV}}&=\frac{2 U^2 e^{-2 \ell E } \left(e^{2 \ell E } \left(3 E^2 \ell ^2+1\right)-2 e^{\ell E }+1\right)}{3 c^2 \ell E ^2}\approx \frac{8 E U^2}{3 c^2}-\frac{2 \ell  \left(E^2 U^2\right)}{3 c^2}+\frac{7 E^3 U^2 \ell ^2}{18 c^2},
\end{align}
where we expanded the result to second order in $\ell$ to demonstrate that these quantities reduce to the SR result when $\ell\rightarrow 0$ and that the two scenarios give distinct average energies only at second order in the quantum gravity parameter. In any case, we will work at finite order in $\ell$ in the discussion below. These expressions are necessary to calculate the energy loss given by $b(E)$ in \eqref{eq:b2}.

We also calculate the average square of the energy shift in both scenarios, which is important for calculating the quantity $d(E)$ given by \eqref{eq:d}. They are given by the following simple expressions:

\begin{align}
    &\langle(\Delta E)^2\rangle_{\kappa,\mathrm{b}}^{(2)}=\frac{4 U^2 \sinh ^2(E \ell )}{3 c^2 \ell ^2},\\
   &\langle(\Delta E)^2\rangle_{\mathrm{LIV}}^{(2)}= \frac{4 U^2 \left(1- e^{-E \ell }\right)^2}{3 c^2 \ell ^2}.
\end{align}

Besides these results, we also consider the equation for $\tau_{\mathrm{esc}}$ given by \eqref{eq:tauesc} and the speed of light given by \eqref{eq:ckappa} in the Fokker-Planck equation \eqref{eq:fokker}. This way, we are able to derive the equations that give $N(E)$ in both scenarios.

\subsubsection{Particle spectrum (super and subluminal cases)}

As in the first-order mechanism, we can define a dimensionless quantity $x=\ell E$ to describe the spectrum $N(x)$ and the spectral index, given by \eqref{eq:spec1}, in a simpler form:

\begin{align}
&\left(e^x-1\right)^2 \left(\left(e^x+1\right)^2 N_{\mathrm{b},\kappa,+}^{(2)}{}''(x)+\left(6 e^x+7 e^{2 x}+1\right) N_{\mathrm{b},\kappa,+}^{(2)}{}'(x)\right)\nonumber\\
&+2 \left(-8 e^x-e^{2 x}-8 e^{3 x}+6 e^{4 x}-1\right) N_{\mathrm{b},\kappa,+}^{(2)}{}(x)=0\label{eq:n2dsr}
\end{align}
and
\begin{align}
&\left(e^x-1\right)^2 \left(-N_{LIV,+}^{(2)}{}''(x)\right)+\left(e^{2 x} (3 x-2)+e^x+1\right) N_{LIV,+}^{(2)}{}'(x)+e^x \left(e^x (3 x+2)+4\right) N_{LIV,+}^{(2)}{}(x)=0,\label{eq:n2liv}
\end{align}
where we used the subscript $+$ to denote the superluminal case $\ell>0$.

We also consider the subluminal case where $\ell<0$, which gives:

\begin{align}
    &\left(e^x-1\right)^2 \left(\left(6 e^x+e^{2 x}+7\right) N_{\mathrm{b},\kappa,-}^{(2)}{}'(x)-\left(e^x+1\right)^2 N_{\mathrm{b},\kappa,-}^{(2)}{}''(x)\right)\nonumber\\
    &+2 \left(8 e^x+e^{2 x}+8 e^{3 x}+e^{4 x}-6\right) N_{\mathrm{b},\kappa,-}^{(2)}{}(x)=0
\end{align}
and

 \begin{align}
     &\left(e^x-1\right)^2 N_{LIV,-}^{(2)}{}''(x)+\left(-3 x+e^x+e^{2 x}-2\right) N_{LIV,-}^{(2)}{}'(x)+\left(3 x-4 e^x-2\right) N_{LIV,-}^{(2)}{}(x)=0.
 \end{align}

\begin{figure}[H]
    \centering
    \begin{subfigure}{0.45\textwidth}
        \centering
        \includegraphics[width=\textwidth]{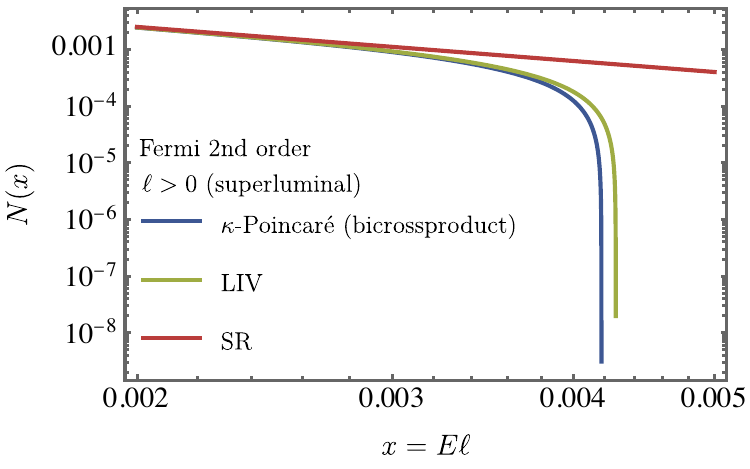}
        \caption{Spectra of Fermi second-order mechanism in the superluminal scenario for DSR (blue), LIV (green), and SR (red) cases.}
        \label{fig:2nd-spectrum-bicross-super}
    \end{subfigure}
    \hfill
    \begin{subfigure}{0.45\textwidth}
        \centering
        \includegraphics[width=\textwidth]{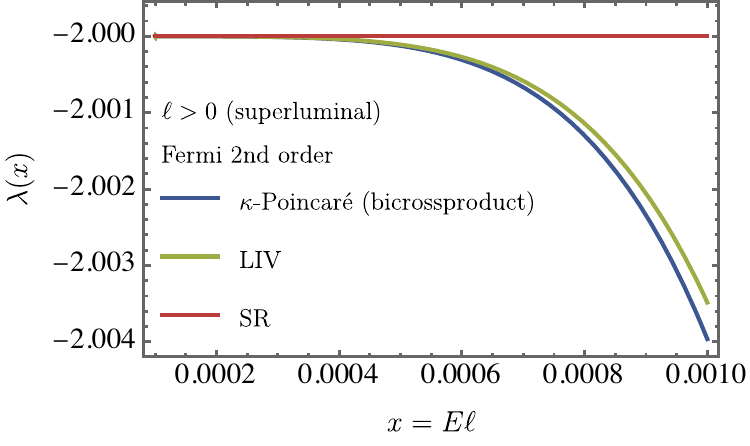}
        \caption{Spectral indexes of Fermi second-order mechanism in the superluminal scenario for DSR (blue), LIV (green), and SR (red) cases.}
        \label{fig:2nd-index-bicross-super}
    \end{subfigure}
    \caption{Second-order Fermi mechanism in the superluminal scenario considering Lorentz preservation, violation, and deformation in the bicrossproduct basis of the $\kappa$-Poincar\'{e} algebra.}
    \label{fig:2nd-bicross-super}
\end{figure}

We could not find analytical solutions for these expressions, but we were able to analyze their behavior. The results are described in Fig.~\ref{fig:2nd-bicross-super}. In Fig.~\ref{fig:2nd-spectrum-bicross-super}, which shows the spectrum in the superluminal scenario, we see a departure from the SR result for energies of the order $(10^{-4}-10^{-3})\ell^{-1}$, suppressing the number of particles compared to the undeformed case. We also notice that LIV introduces stronger effects than the DSR case. This behavior is confirmed by the analysis of the spectral index in Fig.~\ref{fig:2nd-index-bicross-super}.

\begin{figure}[H]
    \centering
    \begin{subfigure}{0.45\textwidth}
        \centering
        \includegraphics[width=\textwidth]{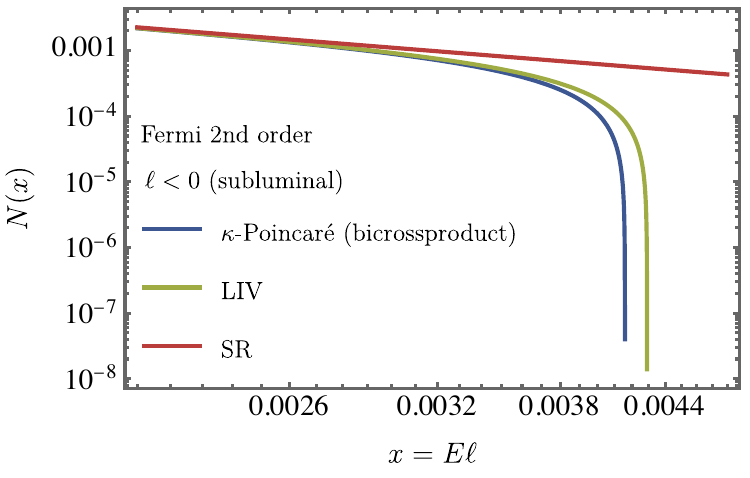}
        \caption{Spectra of Fermi second-order mechanism in the subluminal scenario for DSR (blue), LIV (green), and SR (red) cases.}
        \label{fig:2nd-spectrum-bicross-sub}
    \end{subfigure}
    \hfill
    \begin{subfigure}{0.45\textwidth}
        \centering
        \includegraphics[width=\textwidth]{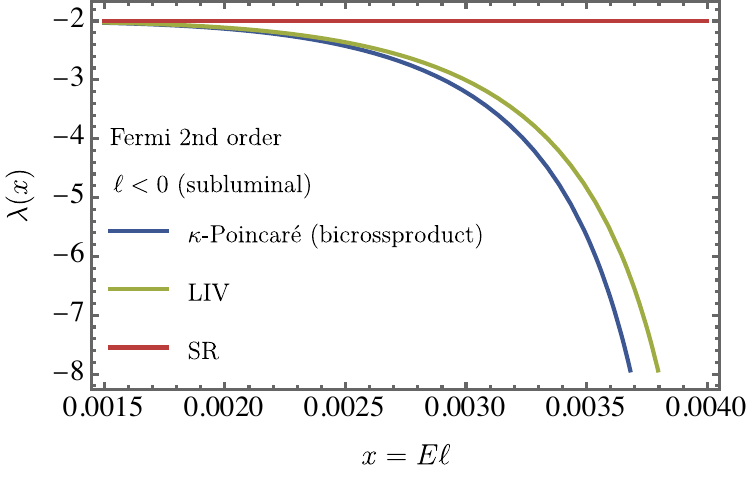}
        \caption{Spectral indexes of Fermi second-order mechanism in the subluminal scenario for DSR (blue), LIV (green), and SR (red) cases.}
        \label{fig:2nd-index-bicross-sub}
    \end{subfigure}
    \caption{Second-order Fermi mechanism in the subluminal scenario considering Lorentz preservation, violation, and deformation in the bicrossproduct basis of the $\kappa$-Poincar\'{e} algebra.}
    \label{fig:2nd-bicross-sub}
\end{figure}

In the subluminal case, in Fig.~\ref{fig:2nd-spectrum-bicross-sub} we observe an initial sudden growth of the spectrum at the energy scale of $10^{-4}\ell^{-1}$, which eventually decreases at the deformation energy scale $\ell^{-1}$. This behavior is also described by the spectral index in Fig.~\ref{fig:2nd-index-bicross-sub}. In this case, the most intense effect is given by the deformed relativistic scenario compared to the Lorentz-violating one.

%%%%%%%%%%%%%%%%%%%%%%%%%%%%%%%%%%%%%%%%%%%%%%%%%%%%%%%%%%%%%%%%%%%%%%%%%%%%%%%%%%%%%%%%%%%%%%%%%

\subsection{Classical basis}

In this section, we analyze the case of undeformed dispersion relations but with deformed energy-momentum conservation. This corresponds to the classical basis of the $\kappa$-Poincar\'{e} algebra as described in section \ref{sec:kappa}. Endowed with the standard Lorentz transformations but a deformed antipode momentum \eqref{eq:om-cl}, we can derive the energy gain:
\begin{equation}
    \Delta E_{\kappa,\mathrm{cl}}^{(2)}=\frac{E U^2 \left((E \ell +1) (E \ell +2)-E \ell  \cos ^2(\theta )\right)}{c^2( E \ell +1)^2}+\frac{ E U (E \ell +1) (E \ell +2) \cos (\theta )}{c( E \ell +1)^2}.
\end{equation}

Since the Lorentz transformations are not modified, the probability of encounter is given by the standard SR result $P(\theta)=1+U\cos(\theta)/c$. Using Eq.~\eqref{eq:aver2}, this gives the average energy:
\begin{align}
    \langle(\Delta E_{\kappa,\mathrm{cl}}^{(2)})\rangle=\frac{E U^2 (E \ell  (4 E \ell +11)+8)}{3 c^2( E \ell +1)^2}\approx \frac{8 E U^2}{3 c^2}-\frac{5 \ell  \left(E^2 U^2\right)}{3 c^2}.
\end{align}

The average of the square of the energy shift is given by:
\begin{equation}
    \langle(\Delta E)^2\rangle_{\kappa,\mathrm{cl}}^{(2)}=\frac{E^2 U^2 (E \ell +2)^2}{3 c^2( E \ell +1)^2}.
\end{equation}

Using these expressions and the results \eqref{eq:b2}, \eqref{eq:d} and \eqref{eq:tauesc} in the Fokker-Planck equation \eqref{eq:fokker}, we can derive the equation for the spectrum $N(E)$ and spectral index according to \eqref{eq:spec1}. Also in this case, we use our defined dimensionless energy given by $x=\ell E$ to derive the following equation:
\begin{align}
&x^2 (x+1) \left((x+1) (x+2)^2 N_{\mathrm{cl}}^{(2)}{}''(x)-2 \left(2 x^2+7 x+7\right) N_{\mathrm{cl}}^{(2)}{}'(x)\right) \nonumber\\
&\quad -2 \left(11 x^4+44 x^3+68 x^2+50 x+12\right) N_{\mathrm{cl}}^{(2)}{}(x)=0.
\end{align}

\begin{figure}[H]
    \centering
    \begin{subfigure}{0.45\textwidth}
        \centering
        \includegraphics[width=\textwidth]{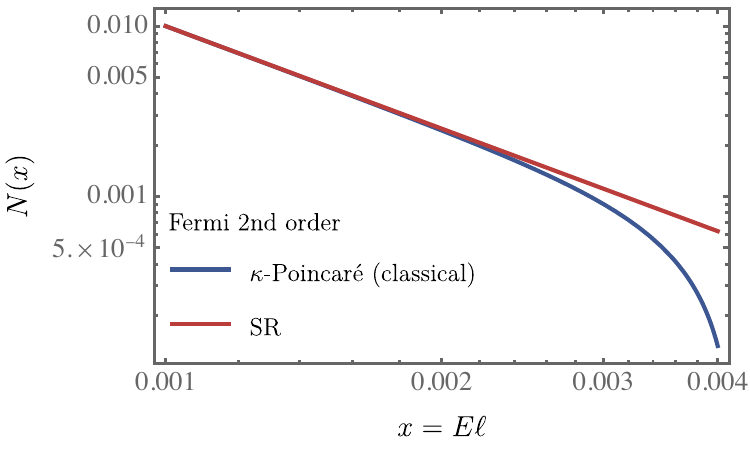}
        \caption{Spectra of Fermi second-order mechanism for DSR (blue) and SR (red) cases.}
        \label{fig:2nd-spectrum-class}
    \end{subfigure}
    \hfill
    \begin{subfigure}{0.45\textwidth}
        \centering
        \includegraphics[width=\textwidth]{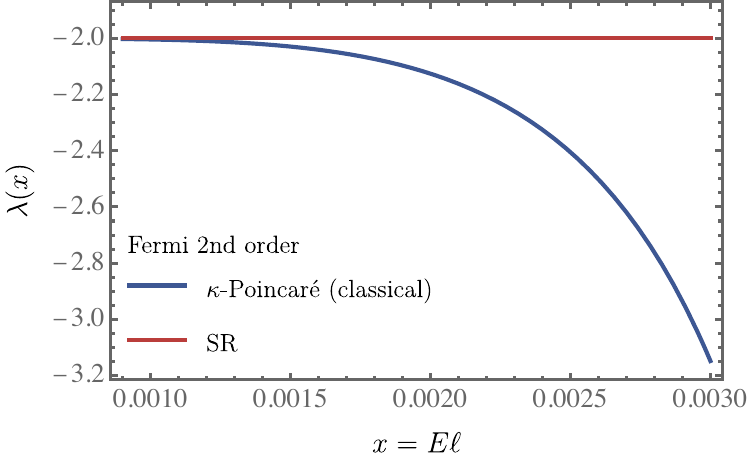}
        \caption{Spectral indexes of Fermi second-order mechanism for DSR (blue) and SR (red) cases.}
        \label{fig:2nd-index-class}
    \end{subfigure}
    \caption{Second-order Fermi mechanism considering Lorentz preservation and deformation in the classical basis of the $\kappa$-Poincar\'{e} algebra.}
    \label{fig:2nd-class}
\end{figure}

From this equation, we verify in Fig.~\ref{fig:2nd-spectrum-class} a decay of the spectrum around energies of the order $10^{-3}\ell^{-1}$ due to the deformed algebra effects, which is confirmed by the spectral index in Fig.~\ref{fig:2nd-index-class}. We observe that the decay of the spectral index is stronger than in the first-order case and does not settle to a specific value. This provides an alternative phenomenological opportunity that can be compared with different data.

\section{Comparison of the predicted source spectra}
\label{sec:compare}

The results of Sections~\ref{sec:first-beyond} and \ref{sec:2nd-beyond} were expressed in terms of the dimensionless variable $x=\ell E$ (or $x=E|\ell|$ in the subluminal case). This is the natural variable in which to exhibit the dependence of the spectra on the deformation or the violation of the symmetry, since in these units the curves become universal and the deformation scale sits simply at $x=1$. In a given physical scenario, however, $|\ell|^{-1}$ is a definite energy, and the spectra become definite functions of $E$. Since the energy is the variable in which cosmic ray observatories report their measurements, and since any future confrontation of these models with data will have to be carried out in this domain, it is instructive to collect in a single place the predictions of all the cases analyzed above, written as functions of $E$. This generalizes to every scenario the illustration presented in Fig.~\ref{fig:dnde} for the classical basis.

All the spectra derived in this work reduce to the special relativistic result $N(E)\propto E^{-2}$ when $E\ll|\ell|^{-1}$, since the deformation is effectively switched off in that regime. It is therefore convenient to normalize them so that they coincide there, which we do by requiring
\begin{equation}\label{eq:norm}
    \lim_{E\to0}E^2N(E)=\mathcal{N},
\end{equation}
with a common constant $\mathcal{N}$ whose value only sets the overall scale of the flux and plays no role in what follows. Any separation between the curves is then due exclusively to the modified kinematics. Introducing the transition energy
\begin{equation}\label{eq:estar-def}
    E_*=|\ell|^{-1},
\end{equation}
the spectra depend on the energy only through the ratio $E/E_*$, so that the figures below, drawn for the illustrative value $E_*=10^{20}\,\mathrm{eV}$, are rigidly translated along the horizontal axis for any other choice of the deformation scale.

\begin{figure}[H]
    \centering
    \begin{subfigure}{0.48\textwidth}
        \centering
        \includegraphics[width=\textwidth]{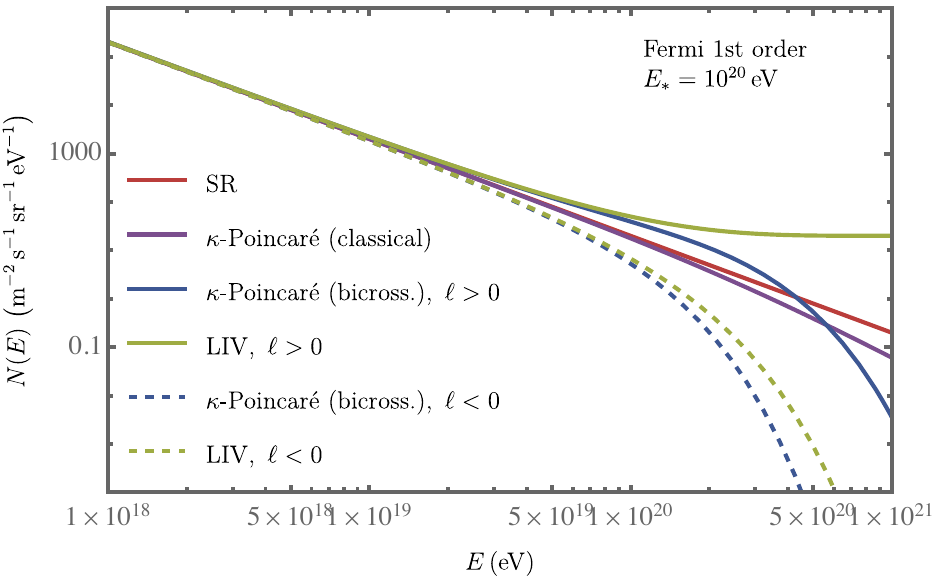}
        \caption{First order.}
        \label{fig:joint-1st}
    \end{subfigure}
    \hfill
    \begin{subfigure}{0.48\textwidth}
        \centering
        \includegraphics[width=\textwidth]{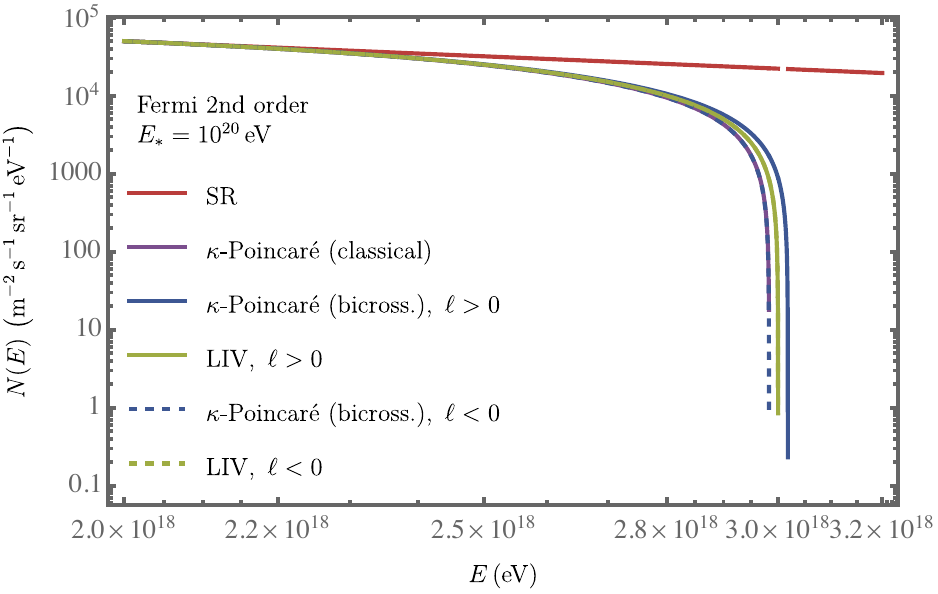}
        \caption{Second order.}
        \label{fig:joint-2nd}
    \end{subfigure}
    \caption{Spectra produced at the source, as functions of the energy, for all the scenarios analyzed in this work: special relativity (red), the classical basis of the $\kappa$-Poincar\'{e} algebra (purple), and the bicrossproduct basis (blue) and the Lorentz-violating case (olive), in the superluminal (solid) and subluminal (dashed) regimes. All the curves are normalized according to \eqref{eq:norm}, so that they coincide at low energy, and the deformation scale is fixed at $E_*=10^{20}\,\mathrm{eV}$.}
    \label{fig:joint-spectra}
\end{figure}

The first-order spectra are collected in Fig.~\ref{fig:joint-1st}. Below $E\sim10^{19}\,\mathrm{eV}$, that is, for $E\lesssim10^{-1}E_*$, all the scenarios are indistinguishable from special relativity, as they must be. Above that energy they separate from one another, and both the direction and the strength of the effect turn out to be strongly scenario dependent; the asymptotic behaviors are collected in Table~\ref{tab:asymptotics}. In the superluminal Lorentz-violating case the spectrum hardens: the spectral index rises from $-2$ and saturates at zero, so that $N(E)$ approaches a constant and the number of particles accelerated to arbitrarily high energies is enhanced with respect to special relativity. This is the only scenario among those considered in which such an enhancement takes place. The superluminal bicrossproduct case behaves differently: the spectrum first hardens slightly, with the index reaching a maximum value $\lambda\simeq-1.61$ around $E\simeq0.8\,E_*$, and beyond that it falls exponentially, $N(E)\propto e^{-E/E_*}$, dropping below the special relativistic curve at about $4\times10^{20}\,\mathrm{eV}$. The two subluminal cases produce instead a monotonic suppression, exponential in both, but with different rates, $N(E)\propto e^{-2E/E_*}$ for the Lorentz-violating case and $N(E)\propto e^{-3E/E_*}$ for the bicrossproduct basis, which is thus the most severely damped of all the scenarios. Finally, the classical basis of the $\kappa$-Poincar\'{e} algebra, in which the dispersion relation and the Lorentz transformations are undeformed and only the composition law is modified, produces the mildest effect: a smooth transition of the spectral index from $-2$ to $-3$, still being $-2.2$ at $E=E_*$ and reaching $-2.8$ only around $10\,E_*$, so that the outcome is a power-law softening rather than a cutoff.

\begin{table}[H]
\centering
\caption{Asymptotic behavior of the first-order source spectra and of the corresponding spectral indices for $E\gg E_*$, with $E_*=|\ell|^{-1}$. In the opposite regime, $E\ll E_*$, every case reduces to $N(E)\propto E^{-2}$ and $\lambda=-2$.}
\label{tab:asymptotics}
\begin{tabular}{lcc}
\hline
Scenario & $N(E)$ for $E\gg E_*$ & $\lambda(E)$ for $E\gg E_*$ \\
\hline
SR & $\propto E^{-2}$ & $-2$ \\
Classical basis & $\propto E^{-3}$ & $-3$ \\
Bicrossproduct, $\ell>0$ & $\propto e^{-E/E_*}$ & $-E/E_*$ \\
LIV, $\ell>0$ & constant & $0$ \\
Bicrossproduct, $\ell<0$ & $\propto e^{-3E/E_*}$ & $-3E/E_*$ \\
LIV, $\ell<0$ & $\propto e^{-2E/E_*}$ & $-2E/E_*$ \\
\hline
\end{tabular}
\end{table}

It is worth stressing that these shapes are, in principle, distinguishable from one another. An exponential suppression, a change of the power-law index by one unit and a hardening towards a flat spectrum are qualitatively different spectral features, and they are tied to different ingredients of the modified kinematics, namely the deformation of the dispersion relation, the modification of the composition law and the modification of the boost transformations. In this sense the spectrum generated at the source is sensitive to a larger portion of the structure of the deformed kinematics than most of the observables usually employed in quantum gravity phenomenology, which are governed by the dispersion relation alone. It is also notable that the effects are not confined to the immediate neighborhood of $E_*$: in most of the cases the departure from the special relativistic behavior is already appreciable at a few tenths of the deformation scale, so that the scale $E_*$ itself need not be reached in order for the acceleration process to carry an imprint of the modified kinematics.

The second-order spectra are shown in Fig.~\ref{fig:joint-2nd}. They are only available numerically, since the corresponding spectral equations admit no closed-form solution, and they are displayed over a much narrower energy interval than the first-order ones. Two features distinguish them from the first-order case. The first is that the scenarios do not separate appreciably from one another before the suppression sets in: the classical basis, the bicrossproduct basis and the Lorentz-violating case, in both regimes, remain close together over the whole interval, so that in this mechanism the shape of the spectrum discriminates much less sharply between the different ways of deforming the kinematics than it does at first order. The second is the position of the suppression itself, which occurs at energies far below the deformation scale, around $3\times10^{18}\,\mathrm{eV}$ for $E_*=10^{20}\,\mathrm{eV}$, that is, at $E\sim10^{-2}E_*$. Taken at face value, this indicates that for a given deformation scale the second-order mechanism departs from the special relativistic prediction considerably earlier than the first-order one, and is in this sense the more sensitive probe of the two, at the price of a spectral shape that is less informative about which deformation is at work.

The curves discussed in this section describe the particles that leave the acceleration site and should be understood as injection spectra. The relation between them and the flux that is actually measured at Earth is taken up in Section~\ref{sec:disc}.

%%%%%%%%%%%%%%%%%%%%%%%%%%%%%%%%%%%%%%%%%%%%%%%%%%%%%%%%%%%%%%%%%%%%%%%%%%%%%%%%%%%%%%%%%%%%%%%%%%%%%%%%%%%%%%%%%%%%%%%%%%%%%%%%%%

\section{Discussion}\label{sec:disc}

Fermi acceleration mechanisms are among the most well-known processes by which particles can acquire high energies in astrophysical environments. The first-order mechanism consists of a systematic series of exchanges of relativistic particles between frames, during which they collide elastically. The second-order mechanism consists of a stochastic process in which particles collide with the environment and gain energy. Both methods rely on the dispersion relation, the transformation of energy and momentum between frames, and the conservation rule of momentum for elastic collisions. Any of these three elements can be modified due to quantum spacetime effects. Modifications of dispersion relations are associated with either Lorentz Invariance Violation (LIV) or Deformed Special Relativity (DSR) scenarios, while deformations of the Lorentz transformations and the composition rule are specific to Deformed Special Relativity.

In this paper, we considered both Fermi mechanisms in a general way that incorporates these ingredients. We also derived expressions for the spectrum and spectral index in these cases and specifically analyzed several classes of scenarios. Regarding deformations of the dispersion relation, we analyzed cases inspired by the bicrossproduct basis of the $\kappa$-Poincar\'{e} algebra, where we considered modifications of the dispersion relation while allowing for the possibility of turning deformations of Lorentz symmetry and the composition law on and off. This allowed us to compare LIV and DSR with Special Relativity (SR) in both superluminal and subluminal scenarios (where massless particles can move with speeds greater or less than $c$). 

Regarding the first-order Fermi mechanism, which could be solved analytically, we found that for the LIV superluminal case, the spectral index stabilizes from $-2$ (the SR result) to zero at very high energies. For the DSR case, the spectral index initially grows but eventually drops to negative values below $-2$. In the subluminal case, the spectral index always becomes smaller than $-2$, dropping more intensely in the DSR case. A possible drawback of this approach is that the modified dispersion relation contributes a first-order effect in the quantum gravity scale, meaning that for Planckian corrections, this effect would only be perceivable for particles with energies close to the Planck energy $\sim 10^{28}\, \text{eV}$. Another possibility is that the deformation scale is smaller than the Planck energy, making it detectable with current observations. However, it would then be necessary to reconcile this with scenarios where the dispersion relations of different particles have been constrained at the Planck scale by various channels \cite{Vasileiou:2013vra,LHAASO:2024lub,HAWC:2019gui,PierreAuger:2022liv}, while cosmic rays could exhibit a modified dispersion relation with a deformation scale far from the Planck scale.

A way to circumvent this difficulty is to consider the classical basis of the $\kappa$-Poincar\'{e} algebra. In this case, the dispersion relation is the same as in special relativity, meaning the Lorentz transformations are also undeformed. However, information about the deformation scale is located in the two-particle sector of the algebra, which implies that the conservation law is deformed. This alone is sufficient to introduce new effects, as an elastic collision of a high-energy particle with an obstacle implies that its opposite momentum is given by its antipode. This approach enabled us to consider a case of deformed relativity without a modified dispersion relation, which can provide new insights in the field.

As a result of the first-order Fermi scenario in the classical basis of $\kappa$-Poincar\'{e}, we find a transition in the spectral index from $-2$ to $-3$ at high energies. Since there are no experimental bounds on this case, we could consider a scenario where new physics emerges at energies around, for instance, $10^{15}\, \text{eV}$, implying a smooth transition in the particle spectrum at this order of magnitude. We do not aim in this paper to explain transitions in cosmic ray spectra, but to demonstrate the power of deformed symmetries in modeling high-energy processes without contradicting previous observations. Similar approaches have recently been explored in \cite{Carmona:2025fdu} regarding gamma rays.

The comparison of the different scenarios carried out in Section~\ref{sec:compare} places these first-order results in a common frame. At the qualitative level, the picture that emerges is clear: the modifications of the kinematics leave distinct and, in principle, distinguishable imprints on the spectrum produced at the source, ranging from a hardening towards a flat spectrum in the superluminal Lorentz-violating case, to a softening of the spectral index by one unit in the classical basis, to exponential suppressions of different strengths in the remaining cases. These effects are also sizable: for a deformation scale $E_*=|\ell|^{-1}$, they modify the spectrum by tens of percent already around $E\sim10^{-1}E_*$ for most of the cases, and by orders of magnitude beyond $E_*$. Depending on the assumptions on the deformation scale and on the scenario, features of this magnitude could plausibly be within reach of the sensitivity of current and forthcoming cosmic ray observatories.

Translating this qualitative statement into a quantitative confrontation with data is, however, a considerably more involved task, and one that we do not attempt here. The spectra derived in this work are injection spectra, whereas the flux measured at Earth results from folding them with the mass composition at the source, the spatial distribution and cosmological evolution of the sources, the energy losses undergone during propagation through the cosmic radiation backgrounds, magnetic deflections, and the response and exposure of the detector. These ingredients reshape the spectrum substantially: the spectral index inferred from the measured flux, in particular, is largely controlled by the superposition of attenuated spectra coming from sources at different distances, rather than by the index with which particles are injected at the source. A proper analysis would therefore require embedding the spectral shapes obtained here into a complete propagation framework, along the lines of the combined fits performed by the collaborations \cite{PierreAuger:2016use,PierreAuger:2023ankle,PierreAuger:2024origin}, and treating the deformation scale on the same footing as the remaining astrophysical parameters. We intend to pursue this program in a subsequent work.

Regarding the second-order mechanism, which could not be solved analytically, we found that in both the superluminal and subluminal scenarios inspired by the bicrossproduct basis of the $\kappa$-Poincar\'{e} algebra, DSR effects are stronger than LIV effects, and both lead to a more pronounced decay of the spectrum compared to special relativity. The same holds for the classical basis of the $\kappa$-Poincar\'{e} algebra. However, in none of these cases does the spectral index stabilize at a particular value. In all cases, the effects are triggered at an energy scale around $10^{-2}\ell^{-1}$, which is an aspect that can have interesting phenomenological impact. 

This paper presents a confrontation between the effects of violation and deformation of Lorentz symmetry due to quantum gravity effects. We have shown that different scenarios can predict different outcomes that can be tested against observations from cosmic ray observatories.

%%%%%%%%%%%%%%%%%%%%%%%%%%%%%%%%%%%%%%%%%%%%%%%%%%%%%%%%%%%%%%%%%%%%%%%%%%%%%%%%%%%%%%%%%%%%%%%%%%%%%%%%%%%%%%%%%%%%%%%%%%%%%%%%%%%%%%%%%%%%%%%%%%%%%%%

\section*{Acknowledgments}

We would like to thank Rafael Alves Batista for the insightful discussions. I. P. L. was partially supported by the National Council for Scientific and Technological Development - CNPq, grant 312547/2023-4. V. B. B. was partially supported by the National Council for Scientific and Technological Development- CNPq, grant No. 307211/2020 7.  A. A. A. F. is supported by Conselho Nacional de Desenvolvimento Cient\'{\i}fico e Tecnol\'{o}gico (CNPq) and Fundação de Apoio à Pesquisa do Estado da Paraíba (FAPESQ), project numbers 150223/2025-0 and 1951/2025. E. A. is funded through an undergraduate scholarship by the Universidade Federal da Paraíba (UFPB) - PIBIC program. E. Otoniel acknowledges support from FUNCAP(BP6-0241-00335.01.00/25)

The authors would like to acknowledge networking support by the COST Action BridgeQG (CA23130), the COST Action RQI (CA23115) and the COST Action FuSe (CA24101) supported by COST (European Cooperation in Science and Technology).

\bibliographystyle{utphys}
\bibliography{ref-fermi}

@article{Gubitosi:2011hgc,
    author = "Gubitosi, Giulia and Mercati, Flavio",
    title = "{Relative Locality in $\kappa$-Poincar{\'e}}",
    eprint = "1106.5710",
    archivePrefix = "arXiv",
    primaryClass = "gr-qc",
    doi = "10.1088/0264-9381/30/14/145002",
    journal = "Class. Quant. Grav.",
    volume = "30",
    pages = "145002",
    year = "2013"
}

@book{Spurio:2018knn,
    author = "Spurio, Maurizio",
    title = "{Probes of Multimessenger Astrophysics. Charged cosmic rays, neutrinos, {\ensuremath{\gamma}}-rays and gravitational waves}",
    doi = "10.1007/978-3-319-96854-4",
    isbn = "978-3-319-96853-7, 978-3-319-96854-4",
    publisher = "Springer",
    series = "Astronomy and Astrophysics Library",
    year = "2018"
}

@article{Lobo:2021yem,
    author = "Lobo, Iarley P. and Pfeifer, Christian and Morais, Pedro H. and Batista, Rafael Alves and Bezerra, Valdir B.",
    title = "{Two-body decays in deformed relativity}",
    eprint = "2112.12172",
    archivePrefix = "arXiv",
    primaryClass = "hep-ph",
    doi = "10.1007/JHEP09(2022)003",
    journal = "JHEP",
    volume = "09",
    pages = "003",
    year = "2022"
}

@article{Amelino-Camelia:2024tdk,
    author = "Amelino-Camelia, Giovanni and Lobo, Iarley P. and Palmisano, Giovanni",
    title = "{Anti-de Sitter momentum space in 3D and 4D quantum gravity}",
    eprint = "2403.16721",
    archivePrefix = "arXiv",
    primaryClass = "gr-qc",
    doi = "10.1088/1361-6382/ad3163",
    journal = "Class. Quant. Grav.",
    volume = "41",
    number = "8",
    pages = "085006",
    year = "2024"
}

@book{longair2011high,
  title={High Energy Astrophysics},
  author={Longair, M.S.},
  isbn={9781139494540},
  year={2011},
  publisher={Cambridge University Press}
}

@ARTICLE{1978MNRAS.182..147B,
       author = "Bell, A. R.",
    title = "{The Acceleration of cosmic rays in shock fronts. I}",
    doi = "10.1093/mnras/182.2.147",
    journal = "Mon. Not. Roy. Astron. Soc.",
    volume = "182",
    pages = "147--156",
    year = "1978"
}

@ARTICLE{1978MNRAS.182..443B,
       author = "Bell, A. R.",
    title = "{The acceleration of cosmic rays in shock fronts. II.}",
     doi = "10.1093/mnras/182.3.443",
    journal = "Mon. Not. Roy. Astron. Soc.",
    volume = "182",
    pages = "443--455",
    year = "1978"
}

@article{Borowiec:2009vb,
    author = "Borowiec, A. and Pachol, A.",
    title = "{Classical basis for kappa-Poincare algebra and doubly special relativity theories}",
    eprint = "0903.5251",
    archivePrefix = "arXiv",
    primaryClass = "hep-th",
    doi = "10.1088/1751-8113/43/4/045203",
    journal = "J. Phys. A",
    volume = "43",
    pages = "045203",
    year = "2010"
}

@article{Carmona:2025fdu,
    author = "Carmona, J. M. and Cort{\'e}s, J. L. and Rescic, F. and Reyes, M. A. and Terzi{\'c}, T.",
    title = "{Photon absorption in a doubly special relativity model with undeformed free propagation and total momentum conservation}",
    eprint = "2503.15203",
    archivePrefix = "arXiv",
    primaryClass = "hep-ph",
    doi = "10.1088/1475-7516/2025/07/066",
    journal = "JCAP",
    volume = "07",
    pages = "066",
    year = "2025"
}

@article{Albuquerque:2023icp,
    author = "Albuquerque, Saulo and Bezerra, Valdir B. and Lobo, Iarley P. and Macedo, Gabriel and Morais, Pedro H. and Rodrigues, Ernesto and Santos, Luis C. N. and Var{\~a}o, Gislaine",
    title = "{Quantum Configuration and Phase Spaces: Finsler and Hamilton Geometries}",
    eprint = "2301.09448",
    archivePrefix = "arXiv",
    primaryClass = "gr-qc",
    doi = "10.3390/physics5010008",
    journal = "Physics",
    volume = "5",
    pages = "90--115",
    year = "2023"
}

@article{Mavromatos:2007xe,
    author = "Mavromatos, Nikolaos E.",
    title = "{Lorentz Invariance Violation from String Theory}",
    eprint = "0708.2250",
    archivePrefix = "arXiv",
    primaryClass = "hep-th",
    doi = "10.22323/1.043.0027",
    journal = "PoS",
    volume = "QG-PH",
    pages = "027",
    year = "2007"
}

@article{Amelino-Camelia:2016gfx,
    author = "Amelino-Camelia, Giovanni and da Silva, Mal{\'u} Maira and Ronco, Michele and Cesarini, Lorenzo and Lecian, Orchidea Maria",
    title = "{Spacetime-noncommutativity regime of Loop Quantum Gravity}",
    eprint = "1605.00497",
    archivePrefix = "arXiv",
    primaryClass = "gr-qc",
    doi = "10.1103/PhysRevD.95.024028",
    journal = "Phys. Rev. D",
    volume = "95",
    number = "2",
    pages = "024028",
    year = "2017"
}

@article{Freidel:2005me,
    author = "Freidel, Laurent and Livine, Etera R.",
    title = "{3D Quantum Gravity and Effective Noncommutative Quantum Field Theory}",
    eprint = "hep-th/0512113",
    archivePrefix = "arXiv",
    doi = "10.1103/PhysRevLett.96.221301",
    journal = "Phys. Rev. Lett.",
    volume = "96",
    pages = "221301",
    year = "2006"
}

@article{Loll:2019rdj,
    author = "Loll, Renate",
    title = "{Quantum Gravity from Causal Dynamical Triangulations: A Review}",
    eprint = "1905.08669",
    archivePrefix = "arXiv",
    primaryClass = "hep-th",
    doi = "10.1088/1361-6382/ab57c7",
    journal = "Class. Quant. Grav.",
    volume = "37",
    number = "1",
    pages = "013002",
    year = "2020"
}

@book{Polchinski:1998rq,
    author = "Polchinski, Joseph",
    title = "{String Theory. Vol. 1: An Introduction to the Bosonic String}",
    publisher = "Cambridge University Press",
    series = "Cambridge Monographs on Mathematical Physics",
    year = "1998",
    doi = "10.1017/CBO9780511816079",
    isbn = "978-0-521-67227-6"
}

@article{Ashtekar:2021kfp,
    author = "Ashtekar, Abhay and Bianchi, Eugenio",
    title = "{A Short Review of Loop Quantum Gravity}",
    eprint = "2104.04394",
    archivePrefix = "arXiv",
    primaryClass = "gr-qc",
    doi = "10.1088/1361-6633/abed91",
    journal = "Rept. Prog. Phys.",
    volume = "84",
    number = "4",
    pages = "042001",
    year = "2021"
}

@article{Horava:2009if,
    author = "Horava, Petr",
    title = "{Spectral Dimension of the Universe in Quantum Gravity at a Lifshitz Point}",
    eprint = "0902.3657",
    archivePrefix = "arXiv",
    primaryClass = "hep-th",
    doi = "10.1103/PhysRevLett.102.161301",
    journal = "Phys. Rev. Lett.",
    volume = "102",
    pages = "161301",
    year = "2009"
}

@article{Mattingly:2005re,
    author = "Mattingly, David",
    title = "{Modern tests of Lorentz invariance}",
    eprint = "gr-qc/0502097",
    archivePrefix = "arXiv",
    doi = "10.12942/lrr-2005-5",
    journal = "Living Rev. Rel.",
    volume = "8",
    pages = "5",
    year = "2005"
}

@article{Amelino-Camelia:2000stu,
    author = "Amelino-Camelia, Giovanni",
    title = "{Relativity in space-times with short distance structure governed by an observer independent (Planckian) length scale}",
    eprint = "gr-qc/0012051",
    archivePrefix = "arXiv",
    doi = "10.1142/S0218271802001330",
    journal = "Int. J. Mod. Phys. D",
    volume = "11",
    pages = "35--60",
    year = "2002"
}

@article{Magueijo:2001cr,
    author = "Magueijo, Joao and Smolin, Lee",
    title = "{Lorentz invariance with an invariant energy scale}",
    eprint = "hep-th/0112090",
    archivePrefix = "arXiv",
    doi = "10.1103/PhysRevLett.88.190403",
    journal = "Phys. Rev. Lett.",
    volume = "88",
    pages = "190403",
    year = "2002"
}

@article{Addazi:2021xuf,
    author = "Addazi, A. and others",
    title = "{Quantum gravity phenomenology at the dawn of the multi-messenger era{\textemdash}A review}",
    eprint = "2111.05659",
    archivePrefix = "arXiv",
    primaryClass = "hep-ph",
    doi = "10.1016/j.ppnp.2022.103948",
    journal = "Prog. Part. Nucl. Phys.",
    volume = "125",
    pages = "103948",
    year = "2022"
}

@article{Amelino-Camelia:2008aez,
    author = "Amelino-Camelia, Giovanni",
    title = "{Quantum-Spacetime Phenomenology}",
    eprint = "0806.0339",
    archivePrefix = "arXiv",
    primaryClass = "gr-qc",
    doi = "10.12942/lrr-2013-5",
    journal = "Living Rev. Rel.",
    volume = "16",
    pages = "5",
    year = "2013"
}

@article{Amelino-Camelia:2001com,
    author = "Amelino-Camelia, Giovanni",
    title = "{Space-time quantum solves three experimental paradoxes}",
    eprint = "gr-qc/0107086",
    archivePrefix = "arXiv",
    doi = "10.1016/S0370-2693(02)01223-6",
    journal = "Phys. Lett. B",
    volume = "528",
    pages = "181--187",
    year = "2002"
}

@article{PierreAuger:2016use,
    author = "Aab, Alexander and others",
    collaboration = "Pierre Auger",
    title = "{Combined fit of spectrum and composition data as measured by the Pierre Auger Observatory}",
    eprint = "1612.07155",
    archivePrefix = "arXiv",
    primaryClass = "astro-ph.HE",
    reportNumber = "FERMILAB-PUB-16-618",
    doi = "10.1088/1475-7516/2017/04/038",
    journal = "JCAP",
    volume = "04",
    pages = "038",
    year = "2017",
    note = "[Erratum: JCAP 03, E02 (2018)]"
}

@article{Barcaroli:2015xda,
    author = "Barcaroli, Leonardo and Brunkhorst, Lukas K. and Gubitosi, Giulia and Loret, Niccol{\'o} and Pfeifer, Christian",
    title = "{Hamilton geometry: Phase space geometry from modified dispersion relations}",
    eprint = "1507.00922",
    archivePrefix = "arXiv",
    primaryClass = "gr-qc",
    doi = "10.1103/PhysRevD.92.084053",
    journal = "Phys. Rev. D",
    volume = "92",
    number = "8",
    pages = "084053",
    year = "2015"
}

@article{Lobo:2020qoa,
    author = "Lobo, Iarley P. and Pfeifer, Christian",
    title = "{Reaching the Planck scale with muon lifetime measurements}",
    eprint = "2011.10069",
    archivePrefix = "arXiv",
    primaryClass = "hep-ph",
    doi = "10.1103/PhysRevD.103.106025",
    journal = "Phys. Rev. D",
    volume = "103",
    number = "10",
    pages = "106025",
    year = "2021"
}

@article{Lobo:2016xzq,
    author = "Lobo, Iarley P. and Loret, Niccol{\'o} and Nettel, Francisco",
    title = "{Investigation of Finsler geometry as a generalization to curved spacetime of Planck-scale-deformed relativity in the de Sitter case}",
    eprint = "1611.04995",
    archivePrefix = "arXiv",
    primaryClass = "gr-qc",
    doi = "10.1103/PhysRevD.95.046015",
    journal = "Phys. Rev. D",
    volume = "95",
    number = "4",
    pages = "046015",
    year = "2017"
}

@article{Girelli:2006fw,
    author = "Girelli, Florian and Liberati, Stefano and Sindoni, Lorenzo",
    title = "{Planck-scale modified dispersion relations and Finsler geometry}",
    eprint = "gr-qc/0611024",
    archivePrefix = "arXiv",
    doi = "10.1103/PhysRevD.75.064015",
    journal = "Phys. Rev. D",
    volume = "75",
    pages = "064015",
    year = "2007"
}

@article{Amelino-Camelia:2014rga,
    author = "Amelino-Camelia, Giovanni and Barcaroli, Leonardo and Gubitosi, Giulia and Liberati, Stefano and Loret, Niccol{\'o}",
    title = "{Realization of doubly special relativistic symmetries in Finsler geometries}",
    eprint = "1407.8143",
    archivePrefix = "arXiv",
    primaryClass = "gr-qc",
    doi = "10.1103/PhysRevD.90.125030",
    journal = "Phys. Rev. D",
    volume = "90",
    number = "12",
    pages = "125030",
    year = "2014"
}

@article{Amelino-Camelia:2025ask,
    author = "Amelino-Camelia, Giovanni and Gubitosi, Giulia and Pellecchia, Pietro and Refuto, Marco and Rosati, Giacomo",
    title = "{DSR-relativistic spacetime picture and the phenomenology of Planck-scale-modified time dilation}",
    eprint = "2506.08111",
    archivePrefix = "arXiv",
    primaryClass = "gr-qc",
    month = "6",
    year = "2025"
}

@article{Pfeifer:2019wus,
    author = "Pfeifer, Christian",
    title = "{Finsler spacetime geometry in Physics}",
    eprint = "1903.10185",
    archivePrefix = "arXiv",
    primaryClass = "gr-qc",
    doi = "10.1142/S0219887819410044",
    journal = "Int. J. Geom. Meth. Mod. Phys.",
    volume = "16",
    number = "supp02",
    pages = "1941004",
    year = "2019"
}

@article{Barcaroli:2017gvg,
    author = "Barcaroli, Leonardo and Brunkhorst, Lukas K. and Gubitosi, Giulia and Loret, Niccol{\'o} and Pfeifer, Christian",
    title = "{Curved spacetimes with local $\kappa$-Poincar{\'e} dispersion relation}",
    eprint = "1703.02058",
    archivePrefix = "arXiv",
    primaryClass = "gr-qc",
    doi = "10.1103/PhysRevD.96.084010",
    journal = "Phys. Rev. D",
    volume = "96",
    number = "8",
    pages = "084010",
    year = "2017"
}

@article{AlvesBatista:2023wqm,
    author = "Alves Batista, R. and others",
    title = "{White paper and roadmap for quantum gravity phenomenology in the multi-messenger era}",
    eprint = "2312.00409",
    archivePrefix = "arXiv",
    primaryClass = "gr-qc",
    doi = "10.1088/1361-6382/ad605a",
    journal = "Class. Quant. Grav.",
    volume = "42",
    number = "3",
    pages = "032001",
    year = "2025"
}

@article{Lukierski:1991pn,
    author = "Lukierski, Jerzy and Ruegg, Henri and Nowicki, Anatol and Tolstoi, Valerii N.",
    title = "{Q deformation of Poincare algebra}",
    reportNumber = "UGVA-DPT-1991-02-710",
    doi = "10.1016/0370-2693(91)90358-W",
    journal = "Phys. Lett. B",
    volume = "264",
    pages = "331--338",
    year = "1991"
}

@article{Lukierski:1993wxa,
    author = "Lukierski, Jerzy and Ruegg, Henri",
    title = "{Quantum kappa Poincare in any dimension}",
    eprint = "hep-th/9310117",
    archivePrefix = "arXiv",
    doi = "10.1016/0370-2693(94)90759-5",
    journal = "Phys. Lett. B",
    volume = "329",
    pages = "189--194",
    year = "1994"
}

@article{Majid:1994cy,
    author = "Majid, S. and Ruegg, H.",
    title = "{Bicrossproduct structure of kappa Poincare group and noncommutative geometry}",
    eprint = "hep-th/9405107",
    archivePrefix = "arXiv",
    reportNumber = "DAMTP-94-24, UGVA-DPT-1994-03-844",
    doi = "10.1016/0370-2693(94)90699-8",
    journal = "Phys. Lett. B",
    volume = "334",
    pages = "348--354",
    year = "1994"
}

@book{Arzano:2021scz,
    author = "Arzano, Michele and Kowalski-Glikman, Jerzy",
    title = "{Deformations of Spacetime Symmetries}: {Gravity, Group-Valued Momenta, and Non-Commutative Fields}",
    doi = "10.1007/978-3-662-63097-6",
    isbn = "978-3-662-63095-2, 978-3-662-63097-6",
    series = "Lecture Notes in Physics",
    volume = "986",
    month = "6",
    year = "2021"
}

@phdthesis{Pachol:2011tp,
    author = "Pachol, Anna",
    title = "{$\kappa$-Minkowski spacetime: Mathematical formalism and applications in Planck scale physics}",
    eprint = "1112.5366",
    archivePrefix = "arXiv",
    primaryClass = "math-ph",
    school = "Wroclaw U.",
    year = "2011"
}

@article{Jacobson:2002hd,
    author = "Jacobson, T. and Liberati, S. and Mattingly, D.",
    title = "{Threshold effects and Planck scale Lorentz violation: Combined constraints from high-energy astrophysics}",
    eprint = "hep-ph/0209264",
    archivePrefix = "arXiv",
    doi = "10.1103/PhysRevD.67.124011",
    journal = "Phys. Rev. D",
    volume = "67",
    pages = "124011",
    year = "2003"
}

@article{Boncioli:2015cqa,
    author = "Boncioli, D. and di Matteo, Armando and Salamida, Francesco and Aloisio, Roberto and Blasi, Pasquale and Ghia, Piera Luisa and Grillo, Aurelio and Petrera, Sergio and Pierog, Tanguy",
    title = "{Future prospects of testing Lorentz invariance with UHECRs}",
    eprint = "1509.01046",
    archivePrefix = "arXiv",
    primaryClass = "astro-ph.HE",
    doi = "10.22323/1.236.0521",
    journal = "PoS",
    volume = "ICRC2015",
    pages = "521",
    year = "2016"
}

@article{Diaz:2016dpk,
    author = "Diaz, J. S. and Klinkhamer, F. R. and Risse, M.",
    title = "{Changes in extensive air showers from isotropic Lorentz violation in the photon sector}",
    eprint = "1607.02099",
    archivePrefix = "arXiv",
    primaryClass = "hep-ph",
    reportNumber = "KA-TP-19-2016",
    doi = "10.1103/PhysRevD.94.085025",
    journal = "Phys. Rev. D",
    volume = "94",
    number = "8",
    pages = "085025",
    year = "2016"
}

@article{Klinkhamer:2007ak,
    author = "Klinkhamer, F. R. and Risse, M.",
    title = "{Ultra-high-energy cosmic-ray bounds on nonbirefringent modified-Maxwell theory}",
    eprint = "0709.2502",
    archivePrefix = "arXiv",
    primaryClass = "hep-ph",
    reportNumber = "KA-TP-17-2007",
    doi = "10.1103/PhysRevD.77.016002",
    journal = "Phys. Rev. D",
    volume = "77",
    pages = "016002",
    year = "2008"
}

@article{PierreAuger:2021mve,
    author = "Abreu, Pedro and others",
    collaboration = "Pierre Auger",
    title = "{Constraining Lorentz Invariance Violation using the muon content of extensive air showers measured at the Pierre Auger Observatory}",
    doi = "10.22323/1.395.0340",
    journal = "PoS",
    volume = "ICRC2021",
    pages = "340",
    year = "2021"
}

@article{Duarte:2024aff,
    author = "Duarte, Matheus and de Souza, Vitor",
    title = "{Fermi acceleration under Lorentz invariance violation}",
    eprint = "2407.17254",
    archivePrefix = "arXiv",
    primaryClass = "astro-ph.HE",
    doi = "10.1088/1475-7516/2024/10/029",
    journal = "JCAP",
    volume = "10",
    pages = "029",
    year = "2024"
}

@article{Duarte:2025lnw,
    author = "Duarte, Matheus and de Souza, Vitor",
    title = "{Effects of Lorentz invariance violation on charged particles and photon production in astrophysical sources}",
    eprint = "2507.06766",
    archivePrefix = "arXiv",
    primaryClass = "astro-ph.HE",
    month = "7",
    year = "2025"
}

@article{Fermi:1949ee,
    author = "Fermi, Enrico",
    title = "{On the Origin of the Cosmic Radiation}",
    doi = "10.1103/PhysRev.75.1169",
    journal = "Phys. Rev.",
    volume = "75",
    pages = "1169--1174",
    year = "1949"
}

@article{Vasileiou:2013vra,
    author = "Vasileiou, V. and Jacholkowska, A. and Piron, F. and Bolmont, J. and Couturier, C. and Granot, J. and Stecker, F. W. and Cohen-Tanugi, J. and Longo, F.",
    title = "{Constraints on Lorentz Invariance Violation from Fermi-Large Area Telescope Observations of Gamma-Ray Bursts}",
    eprint = "1305.3463",
    archivePrefix = "arXiv",
    primaryClass = "astro-ph.HE",
    doi = "10.1103/PhysRevD.87.122001",
    journal = "Phys. Rev. D",
    volume = "87",
    number = "12",
    pages = "122001",
    year = "2013"
}

@article{LHAASO:2024lub,
    author = "Cao, Zhen and others",
    collaboration = "LHAASO",
    title = "{Stringent Tests of Lorentz Invariance Violation from LHAASO Observations of GRB 221009A}",
    eprint = "2402.06009",
    archivePrefix = "arXiv",
    primaryClass = "astro-ph.HE",
    doi = "10.1103/PhysRevLett.133.071501",
    journal = "Phys. Rev. Lett.",
    volume = "133",
    number = "7",
    pages = "071501",
    year = "2024"
}

@article{HAWC:2019gui,
    author = "Albert, A. and others",
    collaboration = "HAWC",
    title = "{Constraints on Lorentz Invariance Violation from HAWC Observations of Gamma Rays above 100 TeV}",
    eprint = "1911.08070",
    archivePrefix = "arXiv",
    primaryClass = "astro-ph.HE",
    doi = "10.1103/PhysRevLett.124.131101",
    journal = "Phys. Rev. Lett.",
    volume = "124",
    number = "13",
    pages = "131101",
    year = "2020"
}

@article{Martinez-Huerta:2020cut,
    author = "Mart{\'\i}nez-Huerta, Humberto and Lang, Rodrigo Guedes and de Souza, Vitor",
    title = "{Lorentz Invariance Violation Tests in Astroparticle Physics}",
    doi = "10.3390/sym12081232",
    journal = "Symmetry",
    volume = "12",
    number = "8",
    pages = "1232",
    year = "2020"
}

@article{PierreAuger:2020ehi,
    author = "Aab, A. and others",
    collaboration = "Pierre Auger",
    title = "{Measurement of the cosmic-ray energy spectrum above $2.5\times 10^{18}$ eV using the Pierre Auger Observatory}",
    eprint = "2008.06486",
    archivePrefix = "arXiv",
    primaryClass = "astro-ph.HE",
    doi = "10.1103/PhysRevD.102.062005",
    journal = "Phys. Rev. D",
    volume = "102",
    number = "6",
    pages = "062005",
    year = "2020"
}

@article{PierreAuger:2022liv,
    author = "Abreu, Pedro and others",
    collaboration = "Pierre Auger",
    title = "{Testing effects of Lorentz invariance violation in the propagation of astroparticles with the Pierre Auger Observatory}",
    eprint = "2112.06773",
    archivePrefix = "arXiv",
    primaryClass = "astro-ph.HE",
    doi = "10.1088/1475-7516/2022/01/023",
    journal = "JCAP",
    volume = "01",
    pages = "023",
    year = "2022"
}

@article{PierreAuger:2023ankle,
    author = "Abdul Halim, A. and others",
    collaboration = "Pierre Auger",
    title = "{Constraining the sources of ultra-high-energy cosmic rays across and above the ankle with the spectrum and composition data measured at the Pierre Auger Observatory}",
    eprint = "2211.02857",
    archivePrefix = "arXiv",
    primaryClass = "astro-ph.HE",
    doi = "10.1088/1475-7516/2023/05/024",
    journal = "JCAP",
    volume = "05",
    pages = "024",
    year = "2023"
}

@article{PierreAuger:2024origin,
    author = "Abdul Halim, A. and others",
    collaboration = "Pierre Auger",
    title = "{Constraining models for the origin of ultra-high-energy cosmic rays with a novel combined analysis of arrival directions, spectrum, and composition data measured at the Pierre Auger Observatory}",
    eprint = "2305.16693",
    archivePrefix = "arXiv",
    primaryClass = "astro-ph.HE",
    doi = "10.1088/1475-7516/2024/01/022",
    journal = "JCAP",
    volume = "01",
    pages = "022",
    year = "2024"
}

\end{document}